\documentclass[twocolumn]{aastex631}
\usepackage{amsmath, comment, afterpage}

\DeclareMathSymbol{\varOmega}{\mathord}{letters}{"0A}
\DeclareMathSymbol{\varSigma}{\mathord}{letters}{"06}
\DeclareMathSymbol{\varPsi}{\mathord}{letters}{"09}

\submitjournal{ApJ}

\newcommand{\vc}[1]{\boldsymbol{#1}}

\defcitealias{ono16}{O16}
\defcitealias{chang23}{C23}

\let\oldalign\align
\let\oldendalign\endalign
\renewenvironment{align} 
  {\linenomathWithnumbers\postdisplaypenalty=0\oldalign}
  {\oldendalign\endlinenomath}

\begin{document}

\title{``Halfway to Rayleigh" and other Insights to the Rossby Wave Instability}

\author[0000-0003-4703-2053]{Eonho Chang}
\affiliation{Graduate Interdisciplinary Program in Applied Mathematics, University of Arizona, Tucson, AZ 85721, USA}
\affiliation{Department of Astronomy and Steward Observatory, University of Arizona, Tucson, AZ 85721, USA}

\author[0000-0002-3644-8726]{Andrew N.\ Youdin}
\affiliation{Department of Astronomy and Steward Observatory, University of Arizona, Tucson, AZ 85721, USA}
\affiliation{Lunar and Planetary Laboratory, University of Arizona, Tucson, AZ 85721, USA}



\begin{abstract}
The Rossby wave instability (RWI) is the fundamental non-axisymmetric radial shear instability in disks.  The RWI can facilitate disk accretion, set the shape of planetary gaps and produce large vortices. It arises from density and/or temperature features, such as radial gaps, bumps or steps.    A general, sufficient condition to trigger the RWI is lacking, which we address by studying the linear RWI in a suite of simplified models, including incompressible and compressible shearing sheets and global, cylindrical disks.  We focus on enthalpy amplitude and width as the fundamental properties of disk features with various shapes.  We find analytic results for the RWI boundary and growth rates across a wide parameter space, in some cases with exact derivations and in others as a description of numerical results.  Features wider than a scale-height generally become unstable about halfway to Rayleigh instability, i.e.\ when the squared epicyclic frequency is about half the Keplerian value, reinforcing our previous finding.  RWI growth rates approximately scale as  enthalpy amplitude to the 1/3 power, with a weak dependence on width, across much of parameter space.  Global disk curvature affects wide planetary gaps, making the outer gap edge more susceptible to the RWI. Our simplified models are barotropic and height-integrated, but the main results should carry over to more complex and realistic scenarios.

\end{abstract}



\section{Introduction}\label{sec:intro}
The Rossby wave instability arises when radial disk structures, such as bumps or gaps, induce strong pressure gradients and non-Keplerian radial shear \citep{lovelace99,li00}.  The RWI can generate large vortices \citep{li01}, for instance at the edges of planetary gaps \citep{de-val-borro07}, which affects planet migration \citep{lin10}.  The RWI also helps transport matter falling onto accretion disks \citep{kuznetsova22}.

Dust is trapped in both RWI-produced vortices and the rings that trigger the RWI, in agreement with the disk structures observed by ALMA \citep{pinilla17}.  The RWI thus constrains observable rings and vortices \citep{chang23}, for instance by regulating  planet-carved gaps \citep{cimerman23}.  Dust trapped in such rings and vortices can trigger planet formation \citep{cy10,drazkowska18,hu18,lyra24}.

These significant consequences arise from simple considerations.  The RWI does not require vertical motions, baroclinicity or cooling, in contrast to the vertical shear instability \citep{nelson13,lin15} and other thermal disk instabilities \citep{lesur23,klahr23}.  The RWI can be triggered by zonal flows arising from these hydrodynamic \citep{manger20}, or magnetohydrodynamic \citep[MHD;][]{jyk09}, instabilities. RWI analyses that include 3D motions \citep{meheut12,lin13},  cooling  \citep{huang22}, dust feedback \citep{liu23} and non-ideal MHD \citep{cui24} are crucial for a complete understanding, and generally find modest corrections to idealized cases. 

Even for simple cases, a general criterion for the onset of the RWI has been elusive. \citet{ono16} found that the RWI was triggered partway between the Lovelace and Rayleigh criteria, for a variety of barotropic disk features.  The Lovelace criterion, equivalent to a vortensity extrema in isentropic disks, is necessary but insufficient for the RWI \citep{lovelace99}.   The Rayleigh criterion gives axisymmetric instability for disks with radially decreasing angular momentum somewhere, i.e.\ negative squared epicyclic frequency, $\kappa^2$.

\citet{chang23} found that disk bumps (barotropic and baroclinic) triggered RWI when $\kappa^2$ was locally reduced to $\sim60\%$ of the Keplerian value.  We colloquially refer to this criterion as ``halfway to Rayleigh" instability.

This work aims to develop a more fundamental understanding of the RWI  boundary  and growth rates, including the ``halfway to Rayleigh" criterion.  We develop scaling relations using the strength and width of disk features.  We start with simplified shearing sheet models and test against global disk models.  This approach is motivated by previous shearing sheet models studying incompressible \citep{lith07} and compressible \citep{vanon16} shear instability, linear Rossby modes \citep{Umurhan2016} and non-linear RWI with cooling \citep{fung21}. 

We present our method for studying the RWI with shearing sheet models in \S\ref{sec:methods}.  Sections \ref{sec:inc_results} and \ref{sec:comp_results} present our results for the incompressible and compressible sheets, respectively.  We compare to global disks in \S\ref{sec:comp_global}.  A suggesting starting point is the summary of our main results in \S\ref{sec:conclusions}.  

\section{Shearing Sheet RWI Models}\label{sec:methods}

\subsection{The Compressible Shearing Sheet}\label{sec:compshearmodel}
The  shearing sheet models a disk patch centered at radius $R_\mathrm{c}$, rotating at the local Keplerian frequency, $\varOmega$, with cartesian $x,y,z$ coordinates oriented radially, azimuthally and vertically.  Vertical averaging gives the equations of motion \citep{gt78b,johnson05}
\begin{subequations}\begin{align}\frac{D\varSigma}{Dt}&=\varSigma\nabla\cdot\vc{v}\label{eq:ccont}\\\left(\frac{D}{Dt}+2\varOmega\hat{z}\times\right)\vc{v}&=3\varOmega^2x\hat{x}-\frac{1}{\varSigma}\nabla P\label{eq:cmom}\\D(P/\varSigma^\gamma)/Dt&=0\label{eq:cen}\end{align}\end{subequations}
for fluid velocity $\vc{v}$, surface density $\varSigma$, and (height-averaged) pressure $P$, with $D/Dt=\partial/\partial t+\vc{v}\cdot\nabla$.  An ideal gas with adiabatic index $\gamma$, adiabatic motions and no self-gravity or viscosity are assumed.

Combinging Eqs.\ (\ref{eq:ccont}, \ref{eq:cmom}),
\begin{align}\label{eq:vortensity}\frac{Dq}{Dt}&=\frac{\nabla\varSigma\times\nabla{P}}{\varSigma^3}\cdot\hat{z}\,,\end{align}
shows that vortensity, $q\equiv(2\varOmega+\hat{z}\cdot\nabla\times\vc{v})/\varSigma$,   is conserved in the absence of baroclinic effects.

We consider an axisymmetric equlibrium with linear perturbations (using $0, 1$ subscripts, repectively) as $\varSigma=\varSigma_0(x)+\varSigma_1,\,P=P_0(x)+P_1,\,\vc{v}=v_0(x)\hat{y}+u_1\hat{x}+v_1\hat{y}$.   Perturbed quantities have a Fourier dependence  $\propto\exp[\imath(k_yy-\omega{t})]$, and $x$-dependent  amplitudes.

The equilibrium orbital motion is 
\begin{align}\label{eq:v0c}v_0&=-\frac{3}{2}\varOmega{x}+\Delta{v}_0=-\frac{3}{2}\varOmega{x}+\frac{1}{2\varOmega}\frac{d\varPi_0}{dx}\end{align}
where $\varPi_0$, the equilibrium enthalpy, $\varPi=\int{d}P/\varSigma$ gives the non-Keplerian motion, $\Delta{v}_0$. 
The equilibrium vortensity, $q_0=\kappa^2/(2\varOmega\varSigma_0)$,
depends on the squared epicyclic frequency: 
\begin{align}\label{eq:epicyclic}\kappa^2&=2\varOmega\left(2\varOmega+\frac{dv_0}{dx}\right)=\varOmega^2+\frac{d^2\varPi_0}{dx^2}\,.\end{align}  

The linear equations of motion for the Fourier amplitudes (given the same symbols as perturbed quantities for simplicity) are
\begin{subequations}\label{eq:complin}\begin{align}-\imath\Delta\omega\varSigma_1&=-\frac{d}{dx}\left(\varSigma_0u_1\right)-\imath k_y\varSigma_0v_1\label{eq:lincont}\\-\imath\Delta\omega{u_1}-2\varOmega v_1&=-\frac{1}{\varSigma_0}\frac{dP_1}{dx}+\frac{\varSigma_1}{\varSigma_0^2}\frac{dP_0}{dx}\label{eq:linu}\\-\imath\Delta\omega{v_1}+\frac{\kappa^2}{2\varOmega}u_1&=-\imath{k_y}\frac{P_1}{\varSigma_0}\label{eq:linv}\\-\imath\Delta\omega\left(\frac{P_1}{P_0}-\gamma\frac{\varSigma_1}{\varSigma_0}\right)&=-u_1\frac{d\ln(P_0/\varSigma_0^\gamma)}{dx}\label{eq:linad}\end{align}\end{subequations}
with Doppler shifted frequency, $\Delta\omega\equiv\omega-v_0(x)k_y$.
We define  a squared sound speed $c_0^2\equiv\gamma{P}_0/\varSigma_0$,  scale-height $H_0\equiv c_0/\varOmega$, (inverse) entropy lengthscale
\begin{align}L_S^{-1}\equiv\frac{1}{\gamma}\frac{d\ln(P_0/\varSigma_0^\gamma)}{dx}\end{align}
and radial buoyancy frequency
\begin{align}N^2&\equiv-\frac{1}{\gamma\varSigma_0}\frac{dP_0}{dx}\frac{d\ln(P_0/\varSigma_0^\gamma)}{dx}=-\frac{c^2}{\gamma{L}_S}\frac{d\ln(P_0)}{dx}\,.\end{align}

Manipulations
yield an ODE for $\Psi\equiv{P}_1/\varSigma_0$, 
\begin{align}\label{eq:LiODE}\Psi''+B(x)\Psi'&=C(x)\Psi,\end{align}
the shearing sheet version of Eq.\ (15) in \citet{li00} with primes for $x$-derivatives, $B\equiv{d}\ln\mathcal{F}/dx$ and
\begin{subequations}\begin{align}\mathcal{F}&\equiv\frac{\varSigma_0\varOmega^2}{\kappa^2+N^2-\Delta\omega^2},\label{eq:F}\\C&\equiv{k_y^2}+\frac{\varSigma_0}{\mathcal{F} H_0^2}+\frac{2\varOmega k_yB}{\Delta\omega}+C_2,\label{eq:c1}\\C_2&\equiv\frac{1-L_S'}{L_S^2}+\frac{B}{L_S}+\frac{4\varOmega{k}_y}{\Delta\omega{L}_S}-\frac{k_y^2N^2}{\Delta\omega^2}.\label{eq:c2}\end{align}\end{subequations}
This work considers isentropic equilibria with $C_2=1/L_S=N^2=0$.  

The corotation resonance at $\Delta\omega=0$ defines a corotation radius, $x_c$, where $\Re[\Delta\omega(x_c)]=0$.  At the Lindblad resonances, where $\Delta\omega^2=\kappa^2+N^2$ and $1/\mathcal{F}=0$, $B$ is singular.

The  Schr\"odinger form of Eq.\  \eqref{eq:LiODE} uses $\Xi=\sqrt{\mathcal{F}} \Psi$ to obtain \citep{ono16}:
\begin{subequations}\label{eq:ODE_and_D}\begin{align}\Xi''&=D(x)\Xi\label{eq:ODE_D}\\D&=\frac{B'}{2}+\frac{B^2}{4}+C\,.\label{eq:D}\end{align}\end{subequations}
 We solve Eq.\ (\ref{eq:LiODE}), since $\Xi$ is singular at Linblad resonances, but  $D$ is a useful effective potential. 

\begin{figure*}\centering\hspace*{-.4cm}\includegraphics[width=1.0\textwidth]{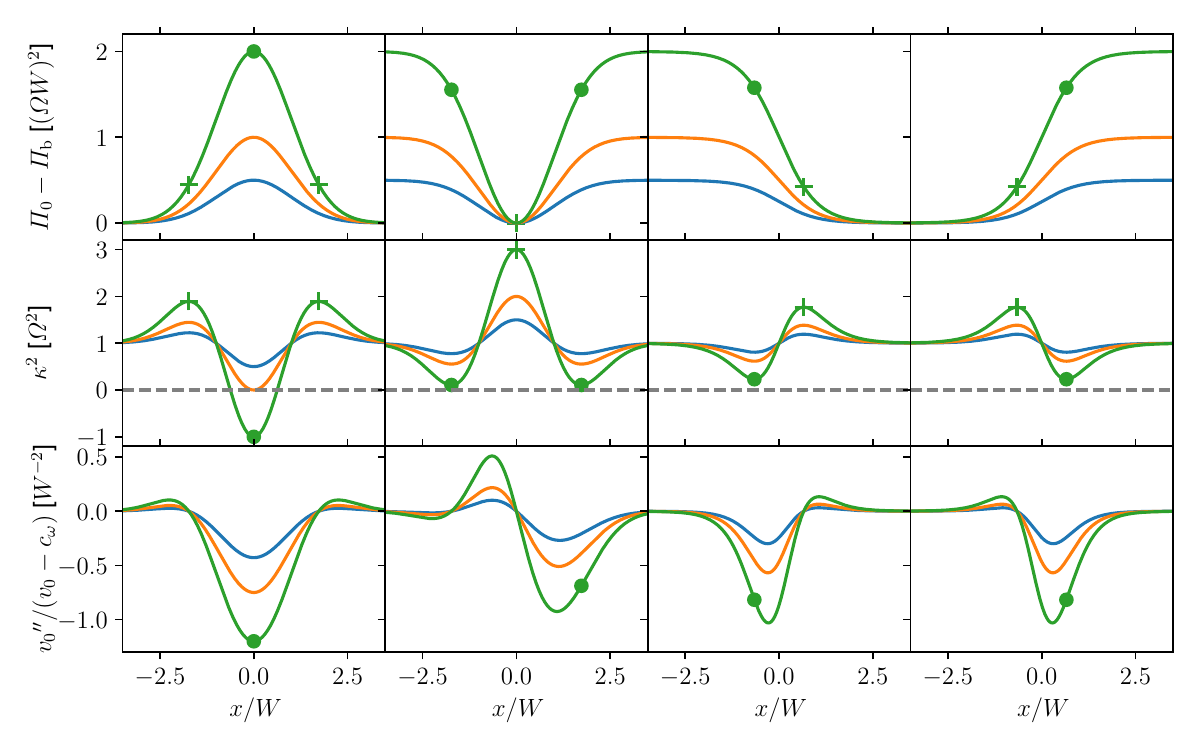}\caption{Radial profiles of shearing sheet equilibria for  (\emph{from left to right:}) bumps, gaps, drops and jumps of width $W$ and amplitude $\mathcal{J}=\Delta\varPi/(\varOmega W)^2=0.5,1,2$ (\emph{blue, orange and green curves}). \emph{Top row:} Enthalpy.  \emph{Middle row}  Epicyclic frequency squared, $\kappa^2$.  Minima (\emph{dots}) and maxima (\emph{pluses}) of vorticity (and equivalently $\kappa^2$) are marked, in all rows. \emph{Bottom row:} The effective potential $D_\mathrm{inc}$ for marginally stable RWI (offset by $k_y^2$, see text).}\label{fig:feats}\end{figure*}

\subsection{The Incompressible Shearing Sheet}
For the incompressible shearing sheet \citep{latter17a} we take the limit $\gamma\rightarrow\infty$, so that 
Eqs.\ (\ref{eq:ccont}, \ref{eq:cen}) give $\nabla\cdot\vc{v}=0$.  We  replace $\nabla{P}/\varSigma=\nabla\varPi$ in equation (\ref{eq:cmom}).  The equilibrium is set by the choice of $\varPi_0(x)$, from which $v_0(x)$ and $\kappa^2$ follow   Eqs.\ (\ref{eq:v0c}, \ref{eq:epicyclic}).  The perturbed flow obeys a stream function, $\psi$, as $u_1=-\imath{k_y}\psi,\,v_1=\psi'$.

The vorticity $\zeta=(\nabla\times\vc{v})\cdot\hat{z}$,  with equilibrium $\zeta_0=v_0'(x)$ and perturbation $\zeta_1=\psi''-k_y^2\psi$ is conserved $D\zeta/Dt=0$.  Thus 
\begin{align}-\imath\Delta\omega\zeta_1&=-\zeta_0'u_1\label{eq:vort}\end{align}
which gives
\begin{align}\psi''=\left(k_y^2+\frac{v_0''}{v_0-\omega/k_y}\right)\psi\equiv{D}_\mathrm{inc}(x)\psi\,,\label{eq:Rayleigh}\end{align}
the famous Rayleigh equation for non-rotating incompressible shear flows.  Coriolis forces set $v_0$, but rotation is otherwise absent \citep[see][]{lith07}.

There is a vast literature on this equation \citep{drazin04}.   Relevant results include Rayleigh's theorem that a vorticity extrema, $\zeta_0'(x)=0$, is required for instability.  Fj\o rtoft's theorem further states that this inflection point must be a maximum in $|\zeta_0(x)|$.  Since disks have $\zeta_0(x)<0$, instability requires a (signed) vorticitiy minimum.  

Fj\o rtoft's theorem agrees with the interpretation of $D_\mathrm{inc}$ as a potential, since for corotation at a vorticity minimum $\Re(D_\mathrm{inc})<0$ near corotation for long wavelengths, $k_y\rightarrow 0$.  We further see that long wavelengths are the most unstable. When applied to compressible, barotropic disks, a \textit{vortensity} minimum is required for instability.

Comparing to the compressible case, we might expect 
$D\rightarrow{D}_\mathrm{inc}$ in some incompressible limit.  Despite a shared $k_y^2$ term, we find that for $k_yW\ll1$, the incompressible limit has $\mathcal{F}\propto1/\kappa^2$ and $2\varOmega{B}\rightarrow-4v_0''$.  Thus the compressbile corotation term is 4 times larger.  This surprising result is possible since $D$ has additional relevant terms and is a potential for a different fluid quantity.  Despite this difference, our compressible results have a well-behaved incompressible limit (\S\ref{sec:comp_results}).

\subsection{Disk Features}\label{sec:shapes}
To understand the universal features of RWI, we consider various disk structures, including bumps, gaps and step.  Our compressible and incompressible models share a common equilibrium enthalpy $\varPi_0(x)$ and thus $v_0(x)$.  Our parameterization
\begin{align}\label{eq:Pibump}\varPi_0(x)={\varPi}_\mathrm{b}+\Delta\varPi{S}(x/W)\end{align} 
has two constants, the reference value ${\varPi}_\mathrm{b}$ (which only affects compressible models) and amplitude $\Delta\varPi>0$.  This work considers the shapes: 
\begin{align}\label{eq:shapes}S(X)=\begin{cases}G(X)&\text{bump}\\1-G(X)&\text{gap}\\\dfrac{1-\tanh(X)}{2}&\text{drop}\\\dfrac{1+\tanh(X)}{2}&\text{jump}\end{cases}\end{align}
for $G(X)=\exp(-X^2/2)$ and scaled distance $X\equiv{x}/W$.    All shapes vary from $0$ to $1$ over a radial width $\sim{W}$.
Since $\varPi_\mathrm{b}=\min(\varPi_0(x))>0$, all  $\varPi_0(x)>0$.

We describe some properties of our shape functions next, then apply them to compressible models in \S\ref{sec:compenth}.

\subsubsection{Shape Functions}  
Figure \ref{fig:feats} plots (in the top row) our enthalpy features. 
The scaled amplitude
\begin{align}\label{eq:J}\mathcal{J}\equiv\Delta\varPi/(\varOmega{W})^2\end{align}
measures a feature's vorticity amplitude.

The middle row of Figure \ref{fig:feats} plots $\kappa^2$.  The location of vorticity (and $\kappa^2$) minima is $x_\mathrm{m}=0$ for bumps, $\pm\sqrt{3}W$ for gaps---which have a pair of vorticity minima---and  $W\ln(2\pm\sqrt{3})/2\simeq\pm1.32W$, for jumps and drops, respectively.  The location of vortensity minima, relevant for compressible flows, will be slightly shifted.

Inner and outer gap edges are symmetrically equivalent in the shearing sheet. So are the drop and jump cases.  Henceforth the ``step'' case refers to both.

Rayleigh instability occurs for $\min(\kappa^2)<0$ and requires vertical motions, absent from our model.  The  Rayleigh instability is still highly relevant and occurs for $\mathcal{J}>\mathcal{J}_\kappa\gtrsim1$.  Specifically, from  
 Equations (\ref{eq:epicyclic}, \ref{eq:Pibump}) 
\begin{align}\label{eq:kappaJ}\frac{\kappa^2}{\varOmega^2}&=1+\mathcal{J}\frac{d^2S}{dX^2}\,.\end{align} 
and $\mathcal{J}_\kappa\equiv1/\max(-d^2S/dX^2)=1$ for bumps, 
2.241 for gaps, and 2.598  for steps.
Thus $\min(\kappa^2)=(1-\mathcal{J}/\mathcal{J}_\kappa)\varOmega^2$ in the shearing sheet.
For global models, $\kappa^2$ also depends on $W/R_\mathrm{c}$ (\S\ref{sec:globalRayleigh}). This  dependence  vanishes in the shearing sheet limit, $W/R_\mathrm{c}\ll1$.

\begin{figure*}\centering\includegraphics[width=\textwidth]{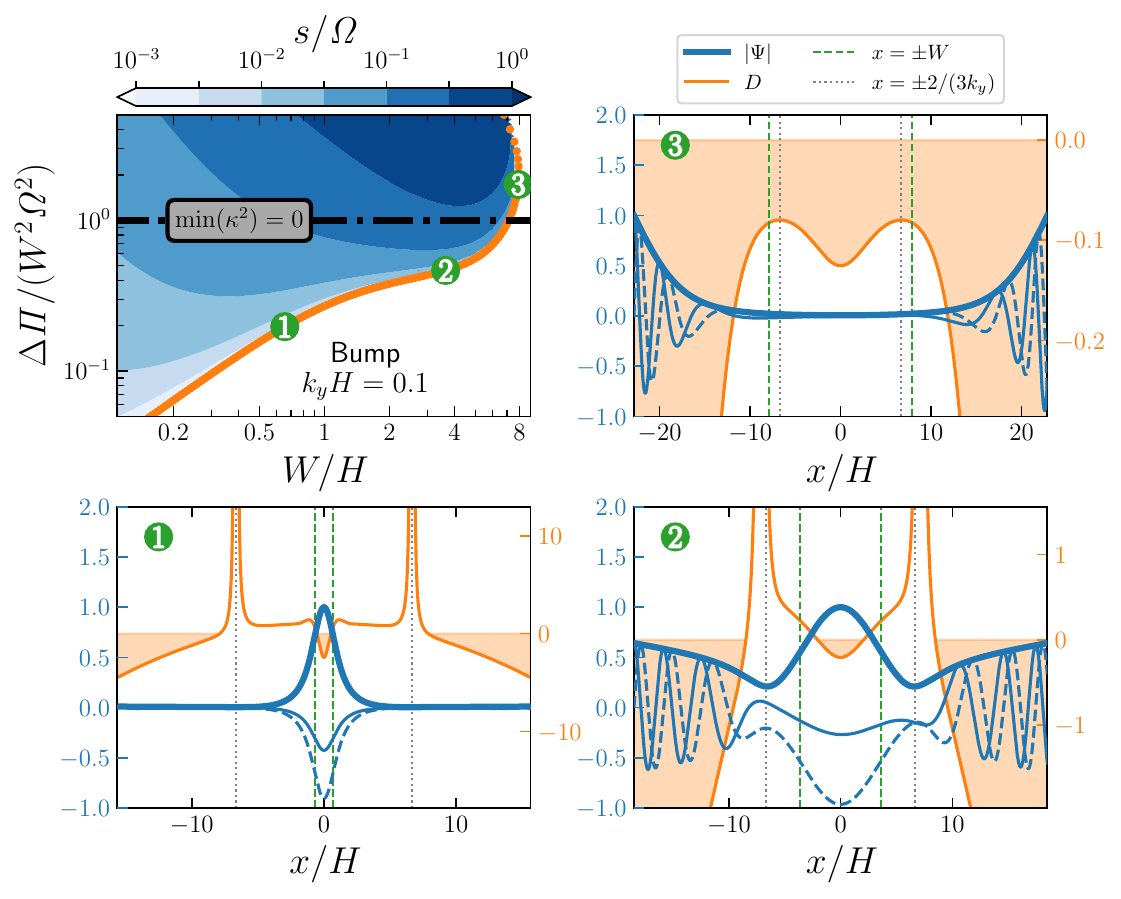}\caption{\emph{Top left:} Along the RWI boundary for $k_yH=0.1$ in a compressible shearing sheet bump, the numbers \textcircled{\raisebox{-0.9pt}{1}}-\textcircled{\raisebox{-0.9pt}{3}} mark the modes investigated. \emph{Other panels:} The effective potential $D$ (\emph{orange curves, with negative regions shaded}) and pressure perturbation $\Psi$ (\emph{blue curves for magnitude [thick],  real and imaginary parts [thin solid and dashed]}) of the numbered modes, with the bump width (\emph{green dotted lines}) and nominal (Keplerian) location of Lindblad resonances  (\emph{gray dotted lines}) marked. \textcircled{\raisebox{-0.9pt}{1}}: Distant Lindblad resonances, with a trapped mode in the Rossby zone.  \textcircled{\raisebox{-0.9pt}{2}}: The trapped Rossby mode couples to density waves exterior to nearby Lindblad resonances. \textcircled{\raisebox{-0.9pt}{3}}: No Lindblad resonances since $\kappa^2<0$, and a ``leaky" potential (negative everywhere).  This Rayleigh unstable region is not our focus.}\label{fig:margstab-rayleigh}\end{figure*}

The bottom row of Figure \ref{fig:feats} plots the incompressible effective potential  as $D_\mathrm{inc}-k_y^2$ (Eq.\  \ref{eq:Rayleigh}).
The corotation radius is at a vorticity minimum, $x_\mathrm{m}$, with phase speed  $c_\omega=\omega/k_y=v_0(x_\mathrm{m})$.  This choice removes the corotation singularity and gives (consistent with Fj\o rtoft's theorem) a negative potential well for trapped modes.  The compressible potential $D$ behaves similarly, but waves also propagate exterior to Lindblad resonances, where $D<0$ (see Fig.\ \ref{fig:margstab-rayleigh}).

\subsubsection{Compressible Shearing Sheet Features}\label{sec:compenth}
The compressible shearing sheet model requires not just $\varPi_0'(x)$ but also $\varSigma_0$ and $P_0$.  We consider polytropic models with $P_0/P_\mathrm{b}=(\varSigma_0/\varSigma_\mathrm{b})^\Gamma$, with reference values $\varSigma_\mathrm{b},\,P_\mathrm{b}$.  The structure index $\Gamma$ could differ from the adiabatic index $\gamma$ (but doesn't here, see below). 

The polytropic enthalpy
\begin{align}\label{eq:enth0}\varPi_0&=\int\frac{dP_0}{\varSigma_0}=\frac{\Gamma}{\Gamma-1}\frac{P_0}{\varSigma_0}\end{align}
 matches Eq.\ \eqref{eq:Pibump} for  
\begin{align}\label{eq:equilibrium_polytropic_sigma}
\varSigma_0&={\varSigma}_\mathrm{b}\left[1+\frac{\Delta\varPi}{{\varPi}_\mathrm{b}}S(x/W)\right]^\frac{1}{\Gamma-1}\,,\end{align}
and ${\varPi}_\mathrm{b}=\Gamma{P}_\mathrm{b}/[(\Gamma-1)\varSigma_\mathrm{b}]$. 

This compressible polytropic model requires 3 additional parameters, besides $k_yW$ and $\mathcal{J}$: 
$\gamma,\,\Gamma$ and $H\equiv{c}/\varOmega$ with 
\begin{align}\label{eq:csquared}{c}^2\equiv\gamma{P}_\mathrm{b}/{\varSigma}_\mathrm{b}=\gamma(\Gamma-1)\varPi_\mathrm{b}/\Gamma\,.\end{align}
We drop $\mathrm{b}$ subscripts from reference $H$ and $c$ values for convenience.  We don't need $\varSigma_\mathrm{b}$ or $P_\mathrm{b}$ independently, as Eq.\ (\ref{eq:LiODE}) only depends on logarithmic derivatives of $\varSigma_0$ and $P_0$.

To reduce parameter space, we fix $\Gamma=\gamma=4/3$ for an adiabatic sheet with $N^2=1/L_S=0$.  A diatomic gas with $\gamma_\mathrm{3D}=7/5$ corresponds to our height integrated  $\gamma=(3\gamma_\mathrm{3D}-1)/(\gamma_\mathrm{3D}+1)=4/3$ \citep{goldreich86, li00}. Thus $H$ is the only additional free parameter our compressible models.

The limits $\Gamma\rightarrow1,\infty$  describe constant temperature and $\varSigma_0$ features, respectively  \citep{chang23}. 
For completeness, the $\Gamma\rightarrow1$ limit of Eq.\ (\ref{eq:equilibrium_polytropic_sigma}) is 
\begin{align}\label{eq:equilibrium_isothermal_sigma}\varSigma_0&=\gamma\frac{P_0}{c^2}={\varSigma}_\mathrm{b}\exp\left[\frac{\gamma\Delta\varPi}{c^2}S\left(\frac{x}{W}\right)\right]\,.\end{align}
with $\varPi_\mathrm{b}(\Gamma-1)\rightarrow{c}^2/\gamma$ remaining finite.   

\subsection{Boundary Conditions and Solution Methods}\label{sec:BCmeth}

Solving our second order ODEs requires a pair of boundary conditions, applied at large distances $|x|\gg{W},\,1/k_y$, and (for the compressible case) $|x|\gg{H}$.

For the incompressible case (Eq.\ \ref{eq:Rayleigh}),  $D_\mathrm{inc}\rightarrow{k}_y^2$  at large $|x|$. 
Physical solutions decay exponentially, with boundary conditions, 
\begin{align}\label{eq:incBC}\psi'=\pm{k}_y\psi\,,\end{align} 
at  large $\mp|x|$.

For the compressible case, boundary conditions exterior to the Lindblad resonances should match onto outgoing density waves.  We seek WKB solutions of the form $\Psi\sim{A}(x)\exp(\imath\int^xk_x(\chi)d\chi)$.  Compared to previous works \citep{li00,ono16,chang23} who used just $k_x$, we find $A(x)$ to lowest order, which improves some numerical results.

First we confirm that outgoing waves have $k_x(x)>0$. 
The large $|x|$, Keplerian limit gives $\Delta\omega\rightarrow3\varOmega{k}_yx/2,\,\mathcal{F}/\varSigma_0\rightarrow-\varOmega^2/\Delta\omega^{2},\,B\rightarrow-3\varOmega{k}_y/\Delta\omega\rightarrow-2/x$ and 
$C\simeq{D}\rightarrow-(\Delta\omega/c_0)^2\rightarrow[3k_yx/(2H_0)]^2$.  To lowest WKB order,  Eq.\ (\ref{eq:LiODE}) gives  $k_x=\pm \sqrt{-C}$, i.e.\ $\Delta\omega^2=(k_xc_0)^2$.  The group velocity
\begin{align}\frac{\partial\omega}{\partial{k}_x}&=\frac{\partial\Delta\omega}{\partial{k}_x}\approx\frac{k_xc^2}{\Delta\omega}\approx\frac{2k_xH_0}{3k_yx}c\end{align} 
confirms that $k_x>0$ for outgoing waves (as $k_y>0$ by convention).

For more accuracy, we adopt the physical optics solution to Eq.\ (\ref{eq:ODE_D}),
\begin{align}\Xi&\sim\frac{c_\Xi}{\sqrt{k_{x,D}}}\exp\left(\imath {\int}^xk_{x,D}(\chi)d\chi\right)\end{align} 
with $k_{x,D}=\sqrt{-D}$ (the desired positive root) and $c_\Xi$ an arbitrary (complex) constant. 

Taking the derivative gives the boundary condition 
\begin{align}\label{eq:XiWKB}\Xi'&=\left(\imath\sqrt{-D}-\frac{1}{4D}\frac{dD}{dx}\right)\Xi\,.\end{align}
The desired boundary condition for $\Psi=\Xi/\sqrt{\mathcal{F}}$  follows as
\begin{align}\label{eq:WKB_BC_Psi}\Psi'&=\left(\imath\sqrt{-D}-\frac{B}{2}-\frac{1}{4D}\frac{dD}{dx}\right)\Psi\,.\end{align} 
At large $|x|$, $d\ln(D)/dx/4\rightarrow1/(2x)$, so that $|\Xi|\propto1/|x|^{1/2}$ and $|\Psi|\propto|x|^{1/2}$, in agreement with our numerical solutions.

Our numerical solutions use the shooting method.  At the inner boundary, $x_i$, we pick an arbitrary $\psi(x_i)$ or $\Psi(x_i)$ and set the derivative with the boundary condition, Eq.\ (\ref{eq:incBC}) or Eq.\  (\ref{eq:WKB_BC_Psi}). We integrate with the Dormand-Prince method (``DOP853'') implemented in \texttt{scipy.integrate.solve\_ivp}. The integrated solution deviates from the outer boundary condition. Using Muller's method, we minimize the residual error and find the complex eigenvalue $\omega\equiv\omega_r+\imath{s}$.
The shooting method requires good initial guesses.  We use known solutions to gradually explore parameter space. 

For global models, we apply the same method but solve Eq.\ (15) in \citet{li00} instead of Eq.\ \eqref{eq:LiODE}.

We have validated our numerical result several ways, including adjusting the outer boundary positions, finding the incompressible limit of compressible results and using different methods for the RWI stability boundary (below).  Similar to \cite{li00}, we derive an energy equation from Eq.\ \eqref{eq:complin}, which after azimuthal averaging (denoted by brackets) is
\begin{align}\label{eq:azim_pert_energy}\frac{\partial}{\partial{t}}\left[\frac{\varSigma_0}{2}\left(\langle|\vc{v}_1|^2\rangle+\frac{\langle\Psi^2\rangle}{c^2}\right)\right]&=\\
-\frac{dv_0}{dx}\varSigma_0\langle{u}_1v_1\rangle-\frac{d}{dx}\langle{P}_1u_1\rangle&\nonumber\end{align} 
We verified that growth rates and eigenfunctions found by our numerical method satisfy this relation, over a range of parameters $\mathcal{J},\,W/H,\,$ and $k_yH$.

\subsubsection{Locating the Stability Boundary}\label{sec:boundmeth}
We find marginally stable modes using a  simplified method \citep{ono16}.  With $s=0$, we fix the corotation radius,  $x_\mathrm{c}$, to vorticity (or vortensity) minima for incompressible (or compressible) models, which sets $\omega=k_y v_0(x_\mathrm{c})$.   With this choice, the ODE has real coefficients, and no corotation singularity, as shown in Figure \ref{fig:feats} for the incompressible case.

One physical parameter, usually $k_yW$ or $W/H$, varies as the shooting parameter (and eigenvalue).  With other parameters held fixed, this method finds marginally stable solutions.  For the incompressible model, this method uses $\psi(x)$ purely real.  
For the compressible models,  $\Psi(x)$ has a complex boundary condition (Eqs. \ref{eq:XiWKB}, \ref{eq:WKB_BC_Psi}).  Thus the eigenvalue ($W/H$) can acquire an imaginary part, which is unphysical.  Usually this imaginary part is negligibly small ($\lesssim10^{-3}$ of the real part), which validates the method. Fig. \ref{fig:margstab-rayleigh} shows example solutions  obtained with this method.  Growth rates away from the stability boundary are also mapped, using the usual method.

For more extreme parameters---near Rayleigh instability and for $k_yW\simeq1$ (placing Lindblad resonances in the Rossby zone)---this method can fail, as it does  (for different reasons) in baroclinic disks \citep{chang23}.  In these cases, we simply measure where $s$ drops to small values.  

It is numerically difficult to find growth rates with $s/\Omega\lesssim10^{-3}$.  Since both methods agree on the stability boundary location (when this simplified method works), the stability boundary is relatively sharp.

\begin{figure*}\centering\includegraphics[width=\textwidth]{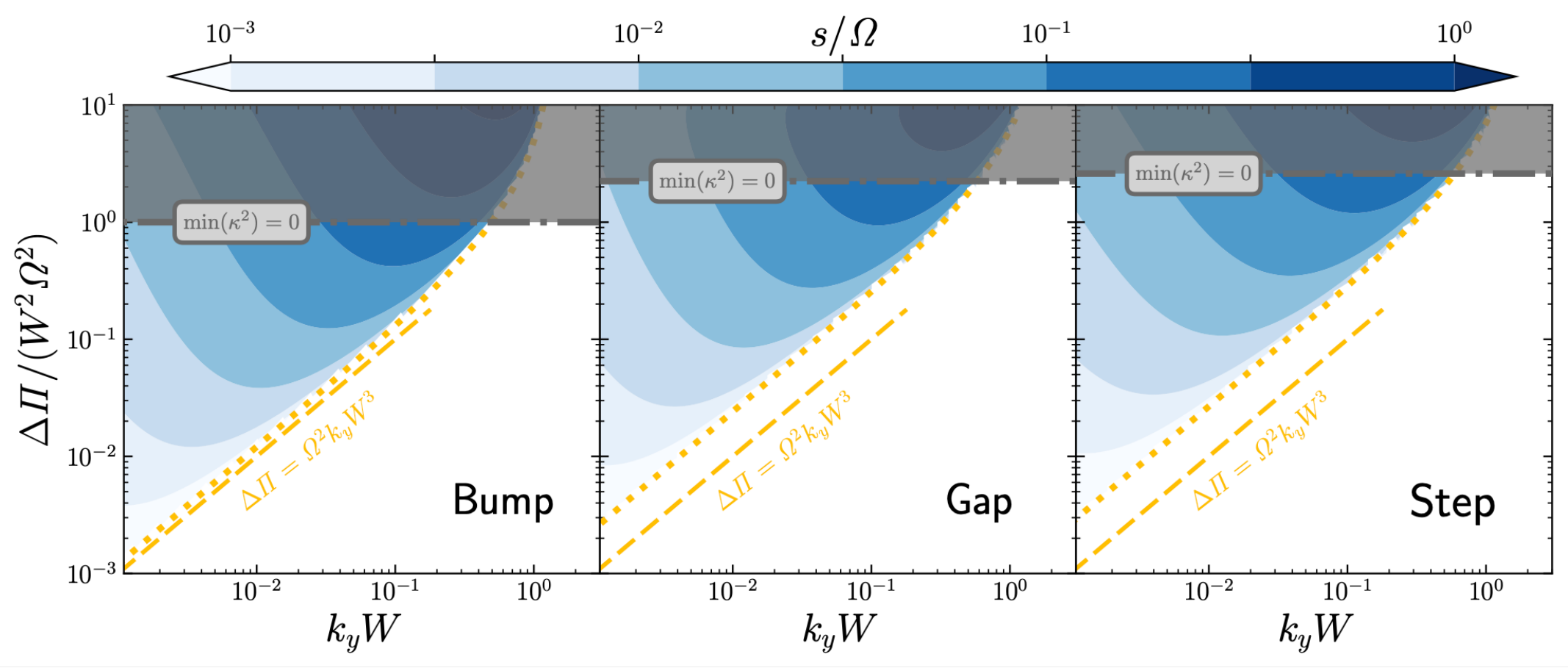}\caption{Incompressible RWI growth rate $s$ for (\textit{left to right}) bump, gap and step features against scaled enthalpy amplitude $\mathcal{J}\equiv \Delta\varPi/(W^2\varOmega^2)$ and the ratio of feature width to azimuthal wavelength, $k_yW$. The RWI boundary (\textit{dotted yellow}),  $\mathcal{J}=k_yW$ reference line (\textit{dashed yellow}), and Rayleigh unstable regions (\textit{gray shaded}) are shown.}\label{fig:ISS-gramp}\end{figure*}

\section{Incompressible Results}\label{sec:inc_results}
Figure \ref{fig:ISS-gramp} maps RWI growth rates for various shapes in the incompressible shearing sheet.  For a given shape, the incompressible RWI is completely described by the parameters for amplitude, $\mathcal{J}=\Delta \varPi/(\varOmega W)^2$, and width (scaled to wavenumber), $k_yW$.
We describe the incompressible stability boundary, growth rates, and eigenfunctions below.

\subsection{Incompressible stability boundary}\label{sec:inc_boundary}
The dotted yellow curves in Figure \ref{fig:ISS-gramp} show the stability boundary, found as described in section \ref{sec:boundmeth}.  RWI occurs for larger $\mathcal{J}$ or smaller $k_yW$ than this boundary, and no modes (stable, unstable or damped) exist on the other side.

The stability boundary is best understood as smoothly connected $\mathcal{J}\ll1$ and ${J}\gg1$ limits.  For $\mathcal{J}\ll1$, the stability boundary follows $\mathcal{J}\simeq{f}_\mathrm{MS}k_yW$, or 
\begin{align}\label{eq:ISS-MS-fit}\Delta\varPi&=f_\mathrm{MS}\varOmega^2k_yW^3\end{align} 
with $f_\mathrm{MS}\simeq1.20,\,2.39,\,2.65$
for the bumps, gaps and steps, respectively.  

For $\mathcal{J}\gg1$, the stability boundary is simply $k_yW=g_\mathrm{MS}$, with $g_\mathrm{MS}=\sqrt{2},1.05,2.0$ for the bump, gap and step cases, respectively.  While large $\mathcal{J}$ values are Rayleigh unstable, this limiting behavior explains why the  stability curve steepens for $\mathcal{J}\gtrsim1$.

These limiting behaviors can be understood several ways, described below.

\begin{figure*}\centering\includegraphics[width=\textwidth]{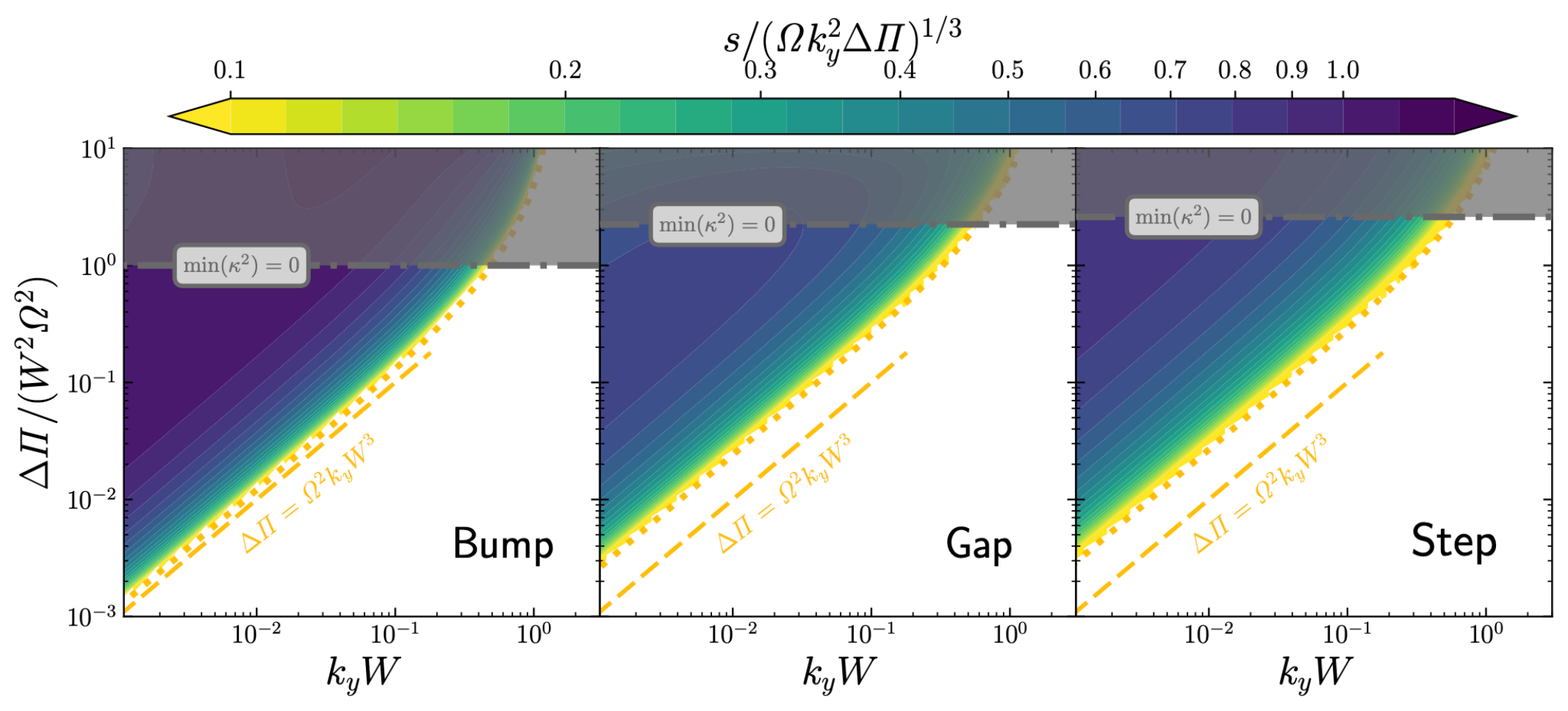}\caption{Similar to Figure \ref{fig:ISS-gramp}, except incompressible growth rates are scaled by a characteristic rate $(\varOmega k_y^2\Delta\varPi)^{1/3}$.  Away from the stability boundary, these scaled rates vary only moderately.}\label{fig:scaled_inc_growth}\end{figure*}

\subsubsection{Intuitive Explanations}\label{sec:inc_intuitive}
The  $\mathcal{J}\gg1$ instability condition, $k_yW<g_\mathrm{MS}$ follows the idea that counter-propagating Rossby waves (CRWs) drive shear instability 
 \citep{heifetz_counter-propagating_1999}.  For simplicity, we consider  bumps and examine the approximate condition for CRWs at $x\simeq\pm{W}$ to maintain stationary phase, with phase speed $c_\omega=\omega/k_y=0$, as illustrated in \S\ref{sec:inc_eigen}.

With a $\psi(x)\propto\exp(\imath{k}_xx)$ WKB approximation,  Eq.\ \eqref{eq:Rayleigh} gives $c_\omega=v_0+v_0''/(k_x^2+k_y^2)$.  At $x=\pm{W}$, $v_0/(\varOmega W)\sim\mp(1+\mathcal{J})$ roughly accounts for Keplerian and non-Keplerian flow, and $v_0''\sim\pm \mathcal{J}\varOmega/W$.  Taking $k_xW\simeq1$ matches the local wave packet to feature size, giving 
\begin{align}\frac{v_0''}{k_x^2+k_y^2}&\sim\pm\frac{\mathcal{J}}{1+(k_yW)^2}(\varOmega{W})\,.\end{align}
Thus $c_\omega=0$ requires $\mathcal{J}\sim(1+\mathcal{J})(1+(k_yW)^2)$.  

For $\mathcal{J}\gg1$ this rough analysis requires $1\sim1+(k_yW)^2$ or $k_yW\lesssim1$ for phase matching and instability, as desired.  For $\mathcal{J}\ll1$, this analysis fails.

Instead, for the $\mathcal{J}\ll1$ boundary, another WKB analysis applies.  Since $k_yW\ll1$, waves have a shallow decay at large $|x|/W$, as $\psi\propto\exp(-k_y|x|)$.  To match onto this decay, the slope across the Rossby zone must change sign, but only change magnitude by a small amount, $\Delta\Phi\equiv{W}\psi'|_{-W}^W/\psi\sim-k_yW$. 

Across corotation, the slope change from WKB oscillations, $\psi\propto\exp(\imath\sqrt{-D_\mathrm{inc}}x)$, is
\begin{align}\label{eq:phaseWKB}\Delta\Phi&=W\int_{-W}^W\psi''dx/\psi\sim{D}_\mathrm{inc}W^2\sim-\mathcal{J},\end{align} 
Where the depth of the potential near corotation $D_\mathrm{inc}\simeq-\mathcal{J}/W^2$ (Fig.\ \ref{fig:feats}).  A trapped mode thus requires $\mathcal{J}\sim{k}_yW$, in agreement with the stability boundary.  The small change in wave phase $\sqrt{-D_\mathrm{inc}} W\sim\sqrt{\mathcal{J}}\ll1$ explains the failure of standard WKB theory for $\mathcal{J}\ll1$, as noted above.

For a more physical explanation of the $\mathcal{J}\ll1$  stability boundary, we briefly summarize the analysis of shearing waves by \citet{lith07}.   Shearing waves interact with  axisymmetric disk features of width $W$ and vorticity amplitude $\Delta\zeta_0\simeq\Delta\Pi/(\varOmega{W}^2)$.

A leading wave with initial radial wavenumber $k_x(t=0)\simeq-1/W$ and fixed $k_y>0$ swings through a radial orientation, $k_x(t_\mathrm{sw})=0$, in time $t_\mathrm{sw}=(2k_x(0)/(3\varOmega{k}_y)\simeq1/(\varOmega{k}_yW)$, since for $\mathcal{J}\ll1$ the flow is nearly Keplerian \citep{glb65II}.  While swinging, the wave couples to the disk feature and spawns a new leading wave.  The amplitude of successive waves increases if $\Delta\zeta_0 t_\mathrm{sw}\gtrsim1$ or $\Delta\varPi\gtrsim\varOmega^2k_yW^3$, reproducing the $\mathcal{J}\ll1$ instability criterion.

\subsubsection{More quantitative explanations}\label{sec:quantexp}
The above arguments can be made more rigorous.  For $\mathcal{J}\ll1$,  \citet{lith07} couples the physical argument ($\Delta \zeta t_\mathrm{sw}\gtrsim1$) to a stability boundary given by the integral
\begin{align}\label{eq:lith_MS}k_y\varOmega&=\frac{1}{3}\int_{-\infty}^{\infty}\frac{d\zeta_0/dx}{x-x_\mathrm{c}}\,dx\end{align} 
with the vorticity minimum at $x_\mathrm{c}$.  This result reproduces Eq.\ (\ref{eq:ISS-MS-fit}), and, for our shapes, precisely gives  $f_\mathrm{MS}=6/I_\mathrm{MS}$ with 
\begin{align}I_\mathrm{MS}=\int_{-\infty}^{\infty}\frac{S'''(X)}{X-X_\mathrm{c}}\,dX\end{align} 
where $X_\mathrm{c}=x_\mathrm{c}/W$.  Integrating $I_\mathrm{MS}$ reproduces our numerical results. For bumps, $f_\mathrm{MS}=3/\sqrt{2\pi}$, and  gaps, $f_\mathrm{MS}=6/\sqrt{2\pi}$. The  numerically integrated $I_\mathrm{MS}$ for steps is also consistent.

For $\mathcal{J}\gg1$, the stability boundary $k_yW=\sqrt{2}$ for the bump case can be derived exactly.   The $\mathcal{J}\rightarrow\infty$ limit gives a parabolic potential well $D_\mathrm{inc}W^2\rightarrow(k_yW)^2+3-(x/W)^2$. This potential has quantized bound states of  ``energy"  $E=3-(k_yW)^2=2n+1$ for $n=0,1,$... \citep[e.g.][Appendix E]{manasreh12}. For $(k_yW)^2>0$ only the $n=0$ bound state exists, which demonstates the lack of RWI modes with higher radial order.  This bound state has $k_yW=\sqrt{2}$, as claimed.  

For all shapes, a necessary condition for RWI follows from the requirement that $D_\mathrm{inc}<0$.  This necessary condition is only close to the stability boundary for $\mathcal{J}\gtrsim1$.  For $\mathcal{J}\gg1$ this necessary condition is $k_yW<\sqrt{3}$ for gaps and bumps and $k_yW<2$ for steps.  These simple necessary conditions are close to, but less strict than, the sufficient conditions for instability.

\begin{figure*}\centering\includegraphics[width=\textwidth]{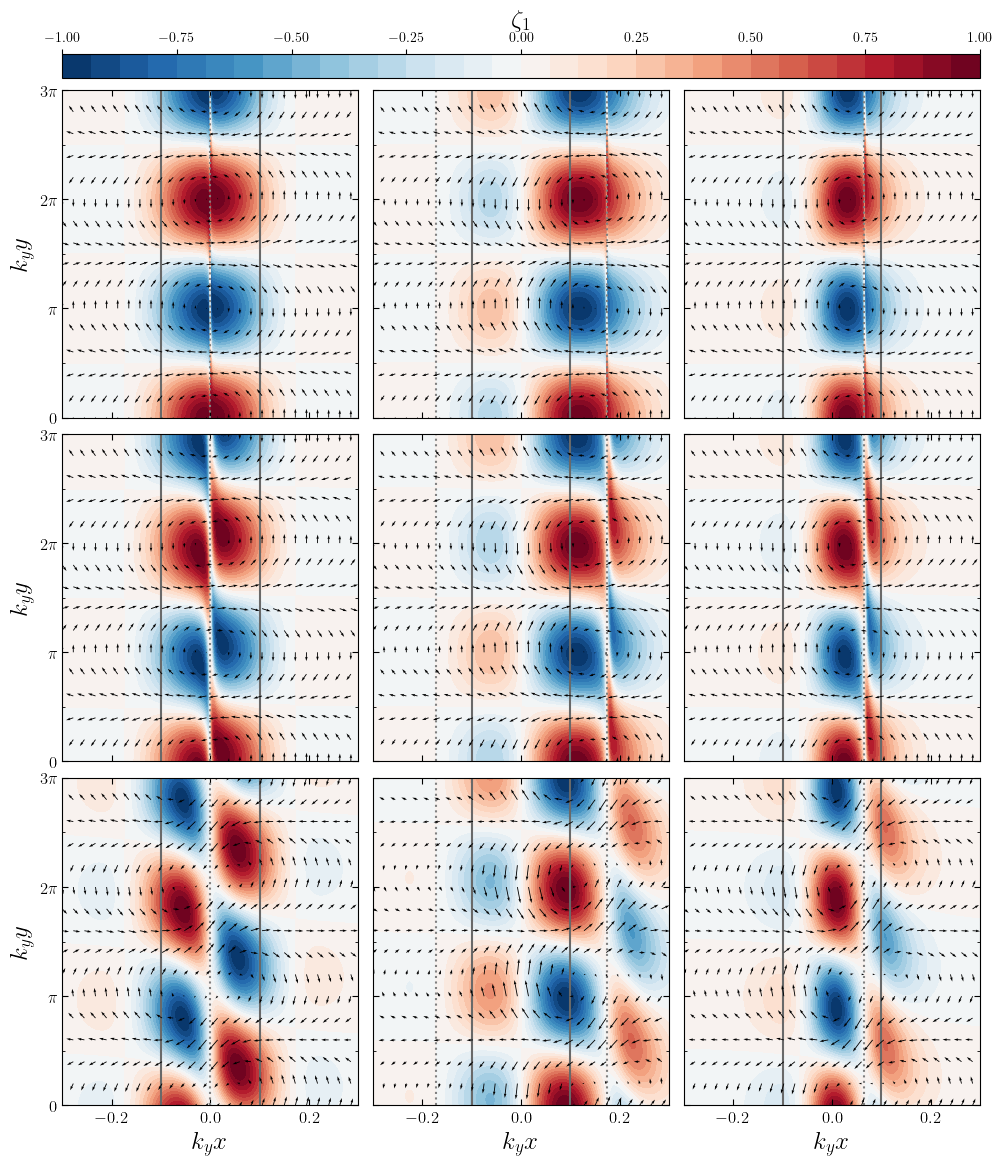}\caption{A map of perturbed vorticity, $\zeta_1$, for the RWI in the incompressible shearing sheet, versus $x,y$ position, with arrows for the the perturbed velocity $\boldsymbol{v}_1$. Rows from \emph{top to bottom} have growth rates $s/\varOmega\approx10^{-3},10^{-2},$ and $10^{-1}$, respectively.  Columns from \emph{left to right} consider enthalpy bumps, gaps and steps, respectively. Solid vertical lines mark the feature width, $x=\pm{W}$. Dotted vertical lines locate the minima of the equilibrium vorticity, $\zeta_0$.   The enthalpy amplitude $\mathcal{J}$ was chosen to produce the desired growth,  with wavenumber $k_y=0.1/W$ in all cases.}\label{fig:ISS-vort-eigfuncs-smallgr}\end{figure*}

\subsection{Incompressible growth rates}
Figure \ref{fig:ISS-gramp} shows that RWI growth rates increase with $\mathcal{J}$ and with $k_yW$, except near marginal stability.  While it is easier to trigger RWI for long wavelengths (small $k_yW$), growth rates well into the unstable region are faster for smaller wavelengths (large $k_yW$).  The smooth variations in growth rates, especially away from the stability boundary, suggest an analytic scaling.

Figure \ref{fig:scaled_inc_growth}  plots growth rates scaled by the characteristic rate
\begin{align}\label{eq:sinc_approx}s_\mathrm{inc}&\equiv(\varOmega{k}_y^2\Delta\varPi)^{1/3}\,.\end{align} 
This approximation is reasonably good, aside from the rapid decay near the stability boundary.

In the $k_yW\ll1$ limit, \citet{lith07} derives the growth rate $s/\varOmega=(3/2)\alpha{k}_yW$, where  $\alpha$ follows from the complex integral constraint on $\alpha$ and $\beta$:
\begin{align}\label{eq:lith_unstable}k_y\varOmega&=\frac{1}{3W}\int_{-\infty}^\infty\frac{d\zeta_0/dX}{(X-\beta)^2+\alpha^2}(X-\beta+\imath\alpha)\,dX.\end{align}
This result reduces to Eq.\ \eqref{eq:lith_MS} for marginal stability, where $\beta\rightarrow{X}_\mathrm{c}$.  Analysis is simplest for bumps, where symmetry about the vorticity minimum gives $\beta=0$, and the imaginary part of the integral vanishes.  

Our parameterization, with $S_\mathrm{B}$ for the bump shape, gives:
\begin{align}\label{eq:alphaint}\frac{k_yW}{\mathcal{J}}&=\frac{1}{f_\mathrm{US}(\alpha)}\equiv\frac{1}{6}\int_{-\infty}^\infty\frac{XS_\mathrm{B}'''(X)}{X^2+\alpha^2}\,dX,\end{align}  
The function $f_\mathrm{US}(\alpha)$ for unstable modes, gives $f_\mathrm{US}(0)=f_\mathrm{MS}$ at marginal stability.  Eq.\ \eqref{eq:alphaint} only gives simple expressions in limiting cases.

For $\alpha\ll1$, $\pi\alpha/2\rightarrow1/f_\mathrm{MS}-k_yW/\mathcal{J}$ gives  the rise in
growth rates near the stability boundary
as
\begin{align}\frac{s}{\varOmega}&\approx\frac{3}{\pi}k_yW\left(\frac{1}{f_\mathrm{MS}}-\frac{k_yW}{\mathcal{J}}\right)\end{align} 

The  $\alpha\rightarrow\infty$ limit gives $f_\mathrm{US}\rightarrow\alpha^4/\sqrt{2\pi}$ and
\begin{align}\label{eq:sinc_lim}\frac{s}{\varOmega}\rightarrow\frac{3}{2}(2\pi)^{1/8}\left(\frac{\Delta\varPi{k}_y^3W}{\varOmega^2}\right)^{1/4}\end{align} 
Unfortunately this limit does not directly apply.  We are mainly interested in $k_yW/\mathcal{J}\gtrsim0.01$, corresponding to $\alpha\lesssim3.4$, i.e.\ at most order unity.  

Our approximate Eq.\ \eqref{eq:sinc_approx}  corresponds to $f_\mathrm{US}\sim\alpha^3$, a good approximation for order unity $\alpha$.

\subsection{Incompressible Eigenfunctions}\label{sec:inc_eigen}

To visualize RWI modes,
Figure \ref{fig:ISS-vort-eigfuncs-smallgr} maps perturbed vorticity  $\zeta_1$ and velocity vectors $\vc{v}_1$ for various growth rates and feature types.  These incompressible eigenfunctions are similar to the standard global, compressible RWI \citep{ono16}.  Corotation is near the vorticity minima marked with the dotted line.

The RWI mechanism is clearest for larger growth rates (bottom row).  The pair of CRWs across corotation (analyzed in \S\ref{sec:inc_intuitive}) is evident.  This wave pair is shifted in azimuthal phase, but radially symmetric for the bumps, and asymmetric for gaps and steps, consistent with their asymmetric potentials (Fig. \ref{fig:feats}).  The aximuthal phase shift causes flow though the vorticity minima to primarily enter regions of negative perturbed vorticity.  This explanation of the growth mechanism is well known for general shear flows \citep{heifetz_counter-propagating_1999} and the  RWI \citep{ono16}. 

At lower growth rates (middle row), the phase shift decreases.  The feeding of negative vorticity from the background into perturbations is less direct, entering narrow fingers near corotation.  Even closer to marginal stability (top rows) the phase shift is nearly gone, and feeding via narrow glitches near corotation is harder to see. 

For marginal stability,  $\zeta_1$ is non-zero and smooth through corototation.  However all growing modes have $\zeta_1=0$ at vorticity extrema (as shown by Eq.\ \ref{eq:vort}).  This fact explains the necessity of small glitches near marginal stability, and the width of prominent CRWs, which fit between a vorticity maximum and minimum (cf.\ Fig. \ref{fig:feats}).

\section{Compressible shearing sheet results}\label{sec:comp_results}

\begin{figure}\centering\includegraphics[width=\columnwidth]{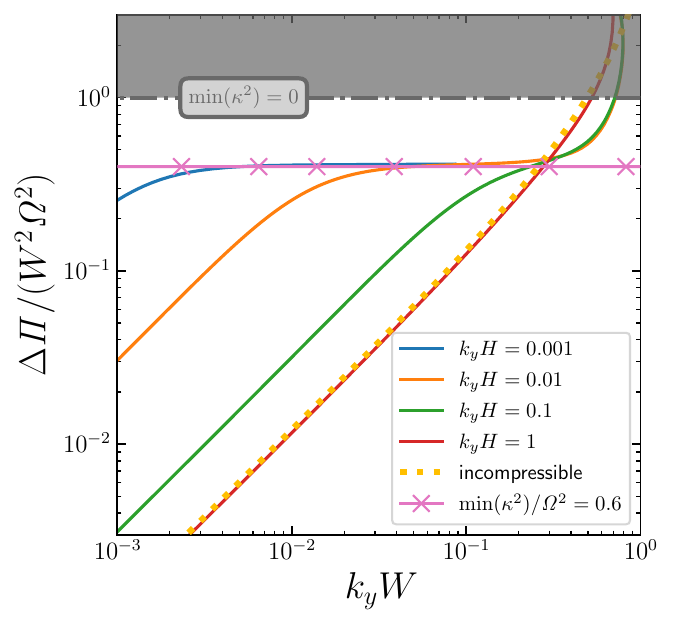}\caption{Marginal stability curves for the RWI of bumps in the compressible shearing sheet, for different values of $k_yH$.   The incompressible, $k_yH\rightarrow\infty$, limit  (\emph{dotted yellow curve}) and $\min(\kappa^2)=0.6$ reference  (\emph{pink line with x's}) are shown.  Axes and Rayleigh unstable region as in Fig. \ref{fig:ISS-gramp}.}\label{fig:effect_compressibility}\end{figure}

We now analyze the compressible shearing sheet model of sections \ref{sec:compshearmodel}, \ref{sec:compenth}. 
Compressible effects are captured by the value of $k_yH$ (see below).  Our incompressible results roughly correspond to the $k_yH\rightarrow\infty$ limit.

For an effective Mach number, we use the Keplerian shear across a lengthscale of $1/k_y$ to define
\begin{align}\mathcal{M}_\mathrm{eff}\equiv\frac{\varOmega/k_y}{c}=\frac{1}{k_yH}\,.\end{align}
RWI modes with $\mathcal{M}_\mathrm{eff} \lesssim1$  behave incompressibly, which is expected of subsonic flows.  In global protoplanetary disks, the RWI is moderately compressible for $m=1$ modes, and more incompressible for higher $m$ (\S\ref{sec:conclusions}, point 3).

\subsection{Compressible Stability Boundary}\label{sec:comp_boundary}

Figure \ref{fig:effect_compressibility} shows the effect of compressibility, measured by $k_yH$, on the RWI boundary.
The bump feature is chosen, and is representative, with quantitative shape effects are noted below.  

For $k_yH=1$ the stability boundary overlaps the incompressible limit ($k_yH\rightarrow\infty$).  As $k_yH$ decreases, compressibility effects increase.

For sufficiently small $k_yH\lesssim0.01$, the stability boundary breaks into three distinct regions,  approximately:
\begin{align}\label{eq:CSS-MS-fit}\frac{\Delta \varPi}{\varOmega^2 W^2}&\approx\begin{cases}f_\mathrm{MS}W\left(k_y+\frac{1}{4H}\right)&\text{if}\,\,W\lesssim{H}\\0.4j_\mathrm{MS}&\text{if}\,\,H\lesssim{W}\lesssim{k}_y^{-1}\\\infty&\text{if}\,\,k_yW\gtrsim{g}_\mathrm{MS}.\\\end{cases}\end{align}
The shape-dependent factors $f_\mathrm{MS},\,g_\mathrm{MS}$ (\S\ref{sec:inc_boundary}) and $j_\mathrm{MS}$ (below) are order unity.  For incompressible parameters, $k_yH\gtrsim0.3$, this stability boundary reverts to the incompressible case, with no intermediate width region. For  marginal compressibility, $k_yH\simeq0.1$, these regions are not as distinct, with overlapping transitions. 

For small widths, $W<H$, the compressible ($k_yH\ll1$) stability boundary follows  $\Delta\varPi\approx{f}_\mathrm{MS}\varOmega^2W^3/(4H)$, independent of $k_y$.  
Compared to the incompressible $k_yW\ll1$ boundary,   $\Delta\varPi\propto{W}^3$ is identical, but compressible enthalpy features must be $\simeq1/(4k_yH)$ larger for instability.  This stabilizing 
effect generally arises from the fact that some of the energy is used to compress the flow \citep{blumen_shear_1970}.   

Wide features and/or short wavelength modes, $k_yW\gtrsim{g}_\mathrm{MS}\sim1$, are RWI-stable, like the incompressible case.  However, compressible modes are more unstable between $0.3\lesssim{k_y}W\lesssim{g}_\mathrm{MS}$ (see Fig.\ \ref{fig:effect_compressibility}).  This effect arises because Lindblad resonances, absent from the incompressible limit, approach the Rossby zone, as described below.  Ultimately, the widest features require a global treatment (\S\ref{sec:comp_global}).

For intermediate widths, with $W\gtrsim{H}$ but $W\lesssim{k}_y^{-1}$, the stability boundary is approximately given by $\mathcal{J}=\Delta\varPi/(W^2\varOmega^2)=0.4j_\mathrm{MS}$ with $j_\mathrm{MS}\simeq1,2.8,3.0$ for bumps, gaps and steps, respectively.  The value of $\min(\kappa^2)=1-\mathcal{J}/\mathcal{J}_\kappa\simeq0.6,0.51,0.54$ for bumps, gaps and steps (respectively) is more similar, emphasizing that $\kappa^2$, and being ``halfway to Rayleigh" is more fundamental.

\begin{figure}\centering\includegraphics[width=\columnwidth]{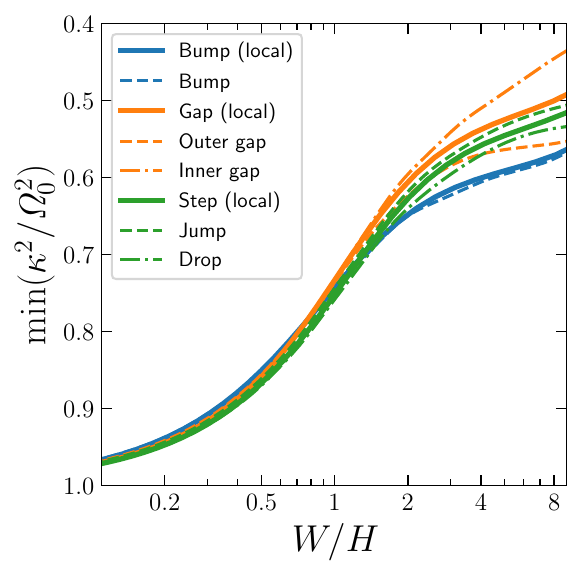}\caption{For marginal RWI, the minimum value of $\kappa^2/\varOmega^2$ caused by bumps, gaps and steps (\emph{blue, orange and green curves}, respectively) vs.\ feature width.  \emph{Solid} curves show compressible shearing sheet models for $k_yH=0.03$. Global models of bumps, outer gap edges and  jumps (\emph{dashed curves}) and of inner gap edges and drops (\emph{dot-dashed curves}) have $m=1,\,h=0.03$ (matching $mh=k_yH$).  Global models break the symmetry between inner and outer gap edges and between jumps and drops.}\label{fig:compressible-kap2min}\end{figure}

Figure \ref{fig:compressible-kap2min} plots $\min(\kappa^2)/\varOmega^2$ for marginal stability, moderate compressibility, $k_yH=0.03$, and different shapes.   (The global models in this figure are discussed in \S\ref{sec:globalminkap}.)  For $W/H\gtrsim2$, the stability boundary is ``halfway to Rayleigh" with $\min(\kappa^2)\simeq0.5{-}0.6\varOmega^2$.  For stronger stronger compressibility (smaller $k_yH$), $\min(\kappa^2)$ values would be more strictly constant  (Figure \ref{fig:effect_compressibility}).

In Figure \ref{fig:compressible-kap2min}, the $k_yW\sim1$ stability boundary is off scale at $W/H\sim30$.  For $W/H\lesssim1$, the stability boundary approaches $\min(\kappa^2)/\varOmega^2=1-(f_\mathrm{MS}/\mathcal{J}_\kappa)(W/H)/4$, following Eq.\ \eqref{eq:CSS-MS-fit}.

While shape effects are minor in the shearing sheet, bumps most readily trigger RWI, at larger $\min(\kappa^2)$ values (Fig. \ref{fig:compressible-kap2min}) and smaller enthalpy amplitudes (smaller $f_\mathrm{MS}$ and $j_\mathrm{MS}$ values, see Fig. \ref{fig:ISS-gramp}).

We next examine origin of the three limits in Eq.\ \eqref{eq:CSS-MS-fit}.

\subsubsection{Small width compressible boundary}
In \S\ref{sec:inc_intuitive},  incompressible instability for $k_yW\ll1$ is given as the \citet{lith07} wave shearing time criteria $\Delta\zeta_0t_\mathrm{sh}\gtrsim1$.  The corresponding $W\ll{H}$ compressible instability criterion is that the sound crossing time $t_\mathrm{sc}\equiv{W}/(\varOmega H)\lesssim\Delta\zeta_0/\varOmega^2$.  Rotation appears explicitly in the compressible (but not the incompressible) instability condition, consistent with the discussion after Eq.\ \eqref{eq:Rayleigh}.

We can also adapt the WKB derivation of the $\mathcal{J},\,k_yW\ll1$ incompressible stability boundary in \S\ref{sec:inc_intuitive} to the compressible $k_yH\ll1,\,W\lesssim{H}$ case.  The main difference in this case is that, from Eq.\ \eqref{eq:c1}, the decay outside the Rossby zone follows $\Psi\propto\exp(-|x|/H)$.  Thus the slope change across corotation (now for $\Psi$) becomes $\Delta \Phi\sim-W/H$.  In this limit, the potential depths are similar (to order unity), so the induced  $\Delta\Phi\sim-\mathcal{J}$.  Matching these two gives $\mathcal{J}\sim{W}/H$, the desired compressible boundary.

\subsubsection{Intermediate width stability boundary}\label{sec:int_width}
The ``halfway to Rayleigh" instability criterion is given above as the scaled enthalpy condition, $\mathcal{J}\gtrsim0.4j_\mathrm{MS}\sim1$.  In absolute terms, relative to $\varPi_{\rm b}\sim c^2$, this condition becomes $\Delta\varPi/c^2\gtrsim(W/H)^2$.  Thus widths larger than $H$ require increasingly strong enthalpy features, a relevant point for the astrophysical origin of these features.

To explain this stability boundary, a negative potential  at corotation $D(x_\mathrm{c})<0$ gives a useful necessary condition for instability (similar to \S\ref{sec:quantexp}).  While simple to state, there are many terms in $D$ to evaluate.  These terms are stabilizing (or destabilizing) if they make a positive (or negative) contribution to $D(x_{\rm c})$.

We focus on the bump case with $x_\mathrm{c}=0$ for simplicity, and take the $k_yW\ll1$ and $\Delta\varPi\gg{c}^2$ (equivalent to $W\gg{H}$ as noted above) limits.  These limits avoid the transitions to neighboring stability regimes.  

The main stabilizing term is $\varSigma_0/(\mathcal{F}H_0^2)$, the usual source of the corotation barrier in disks.  With $\varSigma_0/\mathcal{F}\propto\kappa^2$ this term is reduced near Rayleigh instability, and also, via $H_0$, by disk heating.
In our limits
\begin{align}\label{eq:dw_WH}\left.\frac{\varSigma_0}{\mathcal{F}H_0^2}\right|_{x=0}&\rightarrow\frac{3}{W^2}\left(\frac{1}{\mathcal{J}}-1\right)\end{align}

The main destabilizing term is the corotation term $C_\mathrm{cor}=2\varOmega{k}_yB/\Delta\omega$, though $B'/2$ also contributes.  At corotation $B\propto{d}\ln(q_0)/dx$ diverges approaching Rayleigh instability.

Thus a simple explanation of the ``halfway to Rayleigh" result is that the stabilizing corotation barrier vanishes, and the destabilizing corotation resonance diverges, for $\kappa^2\rightarrow0$.  Thus instability occurs somewhat before this point.

Our limits give
\begin{align}\label{eq:Dcor}C_\mathrm{cor}(0)+\frac{B'(0)}{2}&\rightarrow-\frac{3}{W^2}\left(\frac{3}{2(1-\mathcal{J})}+\frac{1}{3+\mathcal{J}}\right)\end{align} 
with the advertised $\mathcal{J}\rightarrow1$ divergence.

Combining Eqs.\ (\ref{eq:dw_WH}, \ref{eq:Dcor}), the necessary criterion $D(0)<0$ becomes $\mathcal{J}>0.29$.   This condition is close to, but naturally below, the sufficient condition for bumps, $\mathcal{J}>0.4$.

One insight from this analysis is that equation of state effects should have a modest effect on this stability boundary, via $H_0$ and $\varSigma_0'$.  However $\kappa^2$ is the dominant effect.  We defer a more detailed study of thermodynamic, including baroclinic, effects.

\subsubsection{Large width stability boundary}

The $k_yW\gtrsim1$ condition for stability matches the incompressible case which was physically justified in \S\ref{sec:inc_intuitive}.  We do generalize that argument to include compressibility, for reasons explored below.

The enhanced instability of compressible models for  $0.3\lesssim{k_y}W\lesssim1$ is due to the proximity of Lindblad resonances, as noted above.    
While limited to a small region of parameter space, this result does go against the usual trend of compressibility hindering instability.

We expect nearby  Lindblad resonances to enhance RWI because the outer wave propagation zones approach the corotation amplifier \citep{narayan80,tsang08}.  A similar effect is the reduction of the forbidden zone width, i.e.\ Toomre $Q$ barriers, in self-gravitating disks \citep{mark76,gt78b}.

We defer a detailed study of this effect, but note the basic properties of Lindblad resonances in our models.
Their location, where $\Delta \omega^2=\kappa^2$, is at $|x|=\pm2/(3k_y)$ in limit of pure Keplerian flow in the shearing sheet (for corotation at $x=0$). This location clearly approaches $|x|\lesssim{W}$ for $k_yW\gtrsim1$.  Non-Keplerian flow affects the exact location of Lindblad resonances in the Rossby zone.

To understand why Lindblad resonances only affect compressible modes, note that density waves only propagate where $D<0$. 
For Keplerian flow, this propagation region follows from the first two terms in Eq.\ \eqref{eq:c1} as $|x|>(2/3)\sqrt{1/k_y^2+H^2}$, i.e.\ always with $|x|>2H/3$ \citep{pawel93}.  This effective Lindblad resonance location is far from the Rossby zone for incompressible modes with $k_yH\gtrsim1$.

The simplified analyses offered in other regimes is complicated by the presence of Lindblad singularities (where $\mathcal{F}\rightarrow\infty$) in the Rossby zone (see Fig. \ref{fig:margstab-rayleigh}).   
Lindblad singularities can be removed from the ODE  \citep{goldreich86}.  However, they are replaced by ``sonic" singularities at $|x|\simeq2H/3$, which also lie in the Rossby region in this $W>H$ regime.   
We thus defer further analytic exploration of this regime.

\begin{figure*}\centering\includegraphics[width=\textwidth]{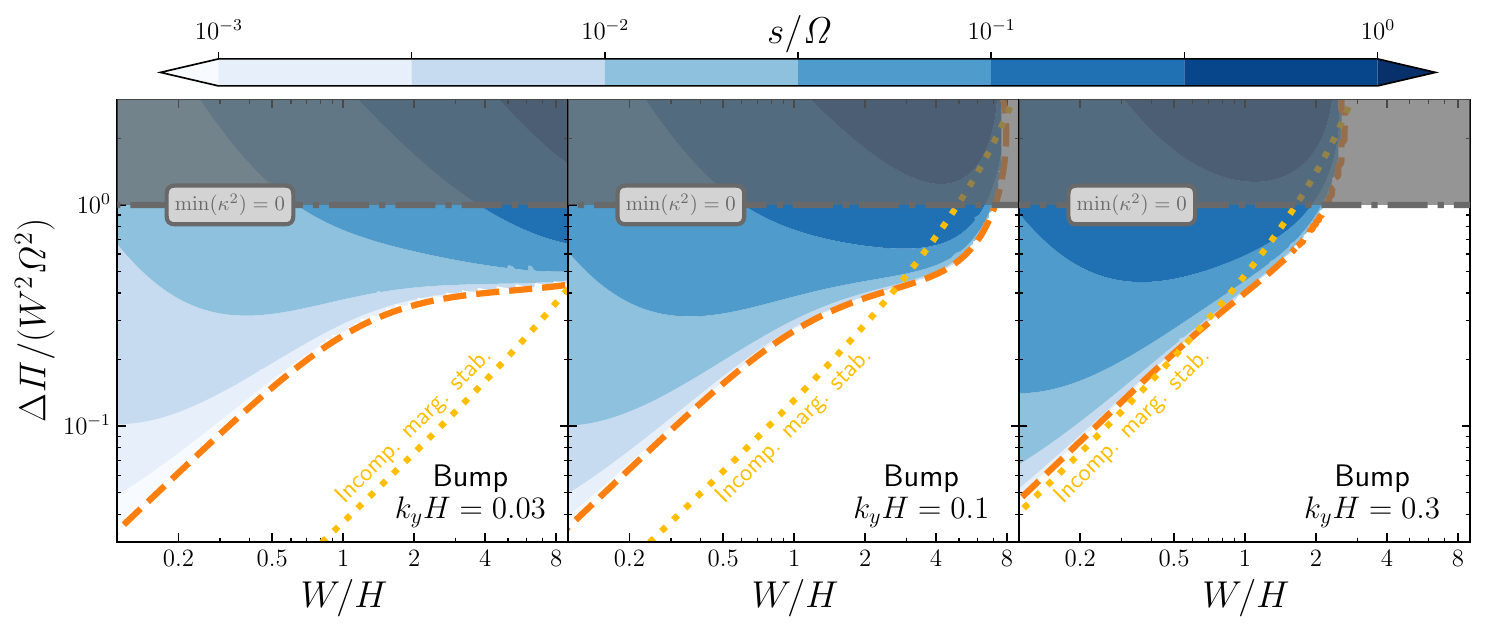}\caption{RWI growth rate, $s$, for bumps in the  compressible shearing sheet, mapped versus scaled bump amplitude, $\Delta\varPi$, and bump width relative to disk scale-height, $W/H$.  Different values of $k_yH$ are shown from left to right.  Compressible (\emph{dashed orange}) and  incompressible (\emph{yellow dotted}) marginal stability curves are compared.}  
\label{fig:CSS-GB}\end{figure*}

\subsection{Compressible growth rates}

Figure \ref{fig:CSS-GB} plots the growth rates for three different levels of compressibility, $k_yH=0.03,0.1,0.3$, from stronger to weaker, which also corresponds to a range of wavelengths from long to short.  The bump case is shown, but other shapes are similar.  The width is plotted as $W/H$, compared to $k_yW$ in Figure \ref{fig:effect_compressibility} for a different perspective.  The characteristic value of $k_yW=0.3$ (where compressible effects transition from stabilizing to destabilizing, as described above) lies at $W/H=10,3,1$ (respectively) in these plots.

For fixed values of $W/H\lesssim1$ the stability boundary is similar for the longer wavelength (more compressible) cases $k_yH=0.03,0.1$ but higher for $k_yH=0.3$ (pushed by the incompressible limit).  However in the unstable region, larger $k_yH$ modes
grow faster.

For larger fixed values of $W/H\gtrsim1$ models with smaller $k_yH$ have an extended region of instability, out to $W/H\sim1/(k_yH)$, but only have faster growth (than larger $k_yH$ modes) close to the stability boundary.  

More quantitatively, we showed that incompressible growth rates are approximately   $s_\mathrm{inc}\sim(\varOmega{k}_y^2\Delta\varPi)^{1/3}$.  For $k_yH\lesssim0.1$, the compressible growth rates in Figure \ref{fig:CSS-GB} are better characterized by a reduced rate: 
\begin{align}\label{eq:scomp_approx}s_\mathrm{comp}&\sim(\varOmega{k}_y^3H\Delta\varPi)^{1/3}\,,\end{align}   
for regions not too close to the stability boundary.  As with the incompressible case, the growth rate has a weak dependence on feature width, $W$, and increases with $k_y$ in the unstable region.

For quick estimates of RWI growth, Eq.\ \eqref{eq:CSS-MS-fit} can be used to determine if parameters are unstable, while the growth rate can be estimated as $\min(s_{\rm inc},s_{\rm comp}$) for parameters  a factor $\gtrsim2$ from the stability boundary.

\section{Comparison to Global Models}
\label{sec:comp_global}

To understand the validity and limitations of our shearing sheet models, we compare to global, cylindrical disk models.  We demonstrate that shearing box models are valid for narrow features with sufficiently small $W/R_\mathrm{c}$, and investigate the role of global disk parameters $m=k_y R_\mathrm{c}$ and $h=H/R_\mathrm{c}$.

We describe our global enthalpy features, and compare to other parameterizations in \S\ref{sec:globalmodels}.  We analyze the Rayleigh stability boundary in global models in \S\ref{sec:globalRayleigh}.  We compare the RWI in shearing sheet and global models in \S\ref{sec:RWIcomparison}, and finally address RWI in dust traps in \S\ref{sec:dusttraps}.

\subsection{Global disk models}\label{sec:globalmodels}

To best compare to our shearing sheet models, our global models use an enthalpy feature
\begin{align}\label{eq:Pi0_global}\varPi_0(R)&=\left(\frac{R}{R_\mathrm{c}}\right)^{q}\left[\varPi_\mathrm{b}+{\Delta\varPi}S(\Delta{R}/W)\right]\,.\end{align}
with $\Delta{R}=R-R_\mathrm{c}$, powerlaw $q$, and using the same shape functions $S$ as Eq.\ \eqref{eq:Pibump}. 

For the same  polytropes,
$P_0\propto\varSigma_0^\Gamma$,
 Eq.\  \eqref{eq:equilibrium_polytropic_sigma} generalizes to 
\begin{align}\label{eq:sigma_global}\varSigma_0(R)&=\varSigma_\mathrm{b}\left(\frac{R}{R_\mathrm{c}}\right)^{n}\left[1+\frac{\Delta\varPi}{\varPi_\mathrm{b}}S(\Delta R/W)\right]^{\tfrac{1}{\Gamma-1}}\end{align} 
with $n=q/(\Gamma-1)$. We again set $\Gamma=\gamma=4/3$ and set $n=q=0$ for simplicity in this work.

Works not using our enthalpy formulation can still be compared to  our results.
Surface density features \citep{ono16,chang23},
\begin{align}\varSigma_{0,\varSigma}(R)&=\varSigma_\mathrm{b}\left(\frac{R}{R_\mathrm{c}}\right)^{n}\left[1+\frac{\Delta \varSigma}{\varSigma_\mathrm{b}}S_\varSigma(\Delta R,W)\right]\,.\end{align} 
are equivalent to our enthalpy formulation, with $\varSigma_0\simeq\varSigma_{0,\varSigma}$, for
\begin{align}\label{eq:sigenth}
\frac{\Delta\varSigma}{\varSigma_\mathrm{b}}S_\varSigma&\simeq\begin{cases}\dfrac{1}{\Gamma-1}\dfrac{\Delta \varPi}{\varPi_\mathrm{b}}S&\text{if}\,\,\dfrac{\Delta \varPi}{\varPi_\mathrm{b}}\ll1,\\\left(\dfrac{\Delta\varPi}{\varPi_\mathrm{b}}S\right)^{\tfrac{1}{\Gamma-1}}&\text{if}\,\,\dfrac{\Delta\varPi}{\varPi_\mathrm{b}}\gg1.\end{cases}\end{align} 
For small amplitudes the shapes are the same and the amplitudes are similar.  Even the isothermal $\Gamma=1$ case follows, from $(\Gamma-1)\varPi_\mathrm{b}\rightarrow{c}^2/\gamma$ (Eq.\ \ref{eq:csquared}).
For larger amplitudes, neither amplitudes nor shapes are the same, so comparisons require more care.

Instead of the above analytic comparison, for general disk features, the corresponding enthalpy profile $\varPi_0(R)$ can be computed, and the amplitude and width measured. For barotropic models, the profile is simply $\varPi_0=\int{d}P_0/\varSigma_0$.
For non-barotropic disks, 
\begin{align}\varPi_0(R)=\varPi_0(R_\mathrm{r})+\int_{R_\mathrm{r}}^R\frac{dP_0/dR'}{\varSigma_0}dR'\,,\end{align} 
for arbitrary values of the reference disk location, $R_\mathrm{r}$, and $\varPi_0(R_\mathrm{r})$.  The reference enthalpy is irrelevant, as a reference sound speed $c$ (which is not arbitrary) can be used instead. Thus, with some effort, the enthalpy properties of any disk feature can be measured, and applied to the results of this work.

\subsection{Rayleigh Instability in Global Models}\label{sec:globalRayleigh}
\begin{figure}\centering\includegraphics[width=\columnwidth]{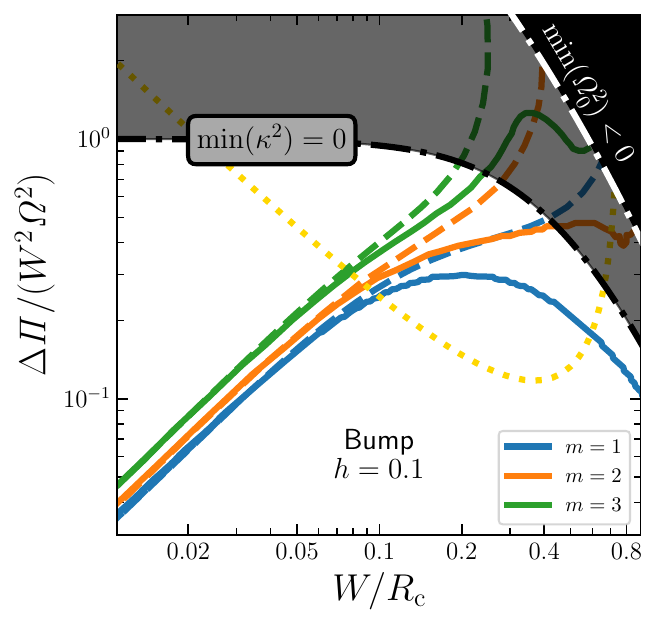}\caption{RWI marginal stability curves for bumps in a $h=0.1$ global disk with azimuthal mode number $m=1,2,3$ (\emph{solid blue, orange, green curves} respectively).  \emph{Dashed curves} show shearing sheet results for comparison.  The greyed out region is Rayleigh unstable.    The \emph{dotted yellow} curve gives the minimum amplitude of dust-trapping bumps.
}\label{fig:global-GB-comparison}\end{figure}

The results for our global models are shown in Fig. \ref{fig:global-GB-comparison} for bumps and Fig. \ref{fig:global-diff-shapes-h0.1} for other shapes.
In global models, the shearing sheet symmetry of inner vs.\ outer gap edges and jumps vs.\ drops is broken.  We show these additional cases.

We first discuss the location of the Rayleigh stability boundary. In shearing sheet models, the Rayleigh stability boundary is at  fixed $\mathcal{J}$ (Eq.\ \ref{eq:kappaJ}, Figs.\ \ref{fig:ISS-gramp}, \ref{fig:effect_compressibility}, \ref{fig:CSS-GB}).
For global models, the critical $\mathcal{J}$ value changes for larger widths, $W/R_\mathrm{c}$.

We describe Rayleigh instability as being ``enhanced" (or ``reduced"), when the critical $\mathcal{J}$ value drops (or increases) for wider features.  Of the shapes considered, most show enhanced Rayleigh instability, to quite different degrees.  Only the inner gap edge shows obviously reduced Rayleigh instability.  The difference with the outer gap edge (which has the strongest enhancement) is striking.  Curiously, the drop shape differs from the inner gap edge, which also can be considered a drop.

We wish to understand these effects, since the Rayleigh stability boundary is crucial for our RWI analysis.  We start with the orbital frequency 
\begin{align}\label{eq:Omega_0}
\varOmega_0^2(R) &= \varOmega_\mathrm{K}^2(R) 
+\frac{1}{R}\frac{d \varPi_0}{dR}  
\end{align}
where the Keplerian $\varOmega_\mathrm{K} = \varOmega_\mathrm{Kc} R_{\rm c}^{3/2}/R^{3/2}$.  Using  Eq.\ \eqref{eq:Pi0_global} with $q =0$ and  $\kappa^2=R^{-3} d(\varOmega_0^2 R^4)/dR$ gives
\begin{align}\label{eq:kappa_global}
\frac{\kappa^2}{\varOmega_\mathrm{Kc}^2}  &= \left(\frac{R_\mathrm{c}}{R}\right)^3+\mathcal{J}\left[ S''(X)
+\frac{3W}{R} S'(X)
\right] 
\end{align} 
for $X=\Delta R/W$, which reduces to the local limit, Eq.\ \eqref{eq:kappaJ},  for $W, \Delta R\ll{R}_\mathrm{c}$. 

There are two main global effects in equation \eqref{eq:kappa_global}.  The first ``Keplerian" effect is the $(R_\mathrm{c}/R)^3$ term, which enhances Rayleigh instability if $R>R_\mathrm{c}$ at $\min(\kappa^2)$.   For example, the outer edge of gaps and jumps have vorticity  (and $\kappa^2$) minima at $R>R_\mathrm{c}$.  Conversely, this effect reduces Rayleigh stability for  inner gap edges and drops.

The second ``non-Keplerian" effect is given by the term $3(W/R)S'(X)$.  
Sub-Keplerian speeds ($S'(X)< 0$) contribute to lower vorticity  and enhanced Rayleigh stability.
This term is  positive (negative) for jumps (drops) and outer (inner) gap edges.  Thus this second effect counteracts the first for these shapes (but not bumps, as discussed last). 

Which effect dominates depends on shape details, especially how far the vorticity (and $\kappa^2$) minimum is from $R_\mathrm{c}$.
For a quantitative criterion, we Taylor expand equation (\ref{eq:kappa_global}) about $W=0$, and evaluate at $R=R_\mathrm{m}=R_\mathrm{c}+X_\mathrm{m}$, the location of the local $\kappa^2$ minimum.  This expansion shows that $\min(\kappa^2)$ is lower, and Rayleigh instability enhanced if 
\begin{align}
f_W &\equiv X_\mathrm{m}-\mathcal{J} S'(X_\mathrm{m})= X_\mathrm{m}+\frac{ S'(X_\mathrm{m})}{S''(X_\mathrm{m})}>0\, .
\end{align} 
The final expression uses $\mathcal{J} =\mathcal{J} _\kappa=-1/S''(X_\mathrm{m})$, the small $W$ Rayleigh stability boundary.  Since $f_W>0$ for outer gap edges (the ``Keplerian" effect dominates) and for drops (the ``non-Keplerian" effect dominates) the enhanced Rayleigh instability of these shapes is explained.  

Shapes with $f_W<0$ (inner gap edges and jumps) are expected to show reduced Rayleigh instability.  This expectation holds for inner gap edges, but jumps are more complicated.   Rayleigh stability is indeed reduced as $W\rightarrow 0$, but this small effect is not visible in Figure \ref{fig:global-diff-shapes-h0.1}.   At larger $W/R_\mathrm{c}$, the first order Taylor expansion is insufficient for jumps.  As $X_\mathrm{m}$ increases for larger $W/R_\mathrm{c}$, the ``Keplerian" effect dominates, explaining the enhanced Rayleigh instability seen for jumps.  The stronger competition between the two effects explains why the jump case shows a weaker enhancement,  starting at larger $W$, compared to other shapes.

The bump case is special with $f_W=0$, which is marginal by our simple criterion.  However since bumps have $S'(X)<0$ for $X>0$, both global effects combine constructively for the bump case, unlike the other cases.  Thus bumps have enhanced Rayleigh stability at larger $W$, with $\min(\kappa^2)$ shifting to $X>0$.

From this analysis, we come to a better understanding of the Rayleigh stability boundary for wide disk features.

For even more extreme parameters than Rayleigh instability, outward pressure gradients can exceed stellar gravity, giving $\varOmega_0^2<0$  (Eq.\ \ref{eq:Omega_0}).  The $\min(\varOmega_0^2)=0$ boundary appears in Figures \ref{fig:global-GB-comparison} and Fig.\ \ref{fig:global-diff-shapes-h0.1}.  This boundary is shown to emphasize how extreme this region of parameter space is.

\subsection{Global RWI vs.\ Shearing Sheet}\label{sec:RWIcomparison}
We now compare the results of our shearing sheet models to the equivalent compressible global models, described in \S\ref{sec:globalmodels}.  Global models introduce the lengthscale, $R_\mathrm{c}$ and thus one additional dimensionless parameter.  Our shearing sheet parameters $\mathcal{J},k_yW$ and $k_yH$ we add the mode number $m=k_yR_\mathrm{c}$.  Removing the local wavenumber $k_y$, an equivalent set---$\mathcal{J},m,h,W/R_\mathrm{c}$---uses the aspect ratio $h=H/R_\mathrm{c}$.

\subsubsection{Effect of Mode Number $m$}\label{sec:m}

Figure \ref{fig:global-GB-comparison} plots the marginal stability curves for $m=1, 2$ and 3 RWI modes in a $h=0.1$ disk versus $W/R_\mathrm{c}$ for a bump shape (other shapes are addressed next).  The equivalent compressible shearing sheet models have $k_yH=mh=0.1, 0.2, 0.3$, and their stability curves are shown for comparison.  

Shearing sheet and global results agree very well for $W/R_\mathrm{c}\lesssim0.1$, as expected.  This agreement is excellent even for the most global $m=1$ modes.  For $W/R_\mathrm{c}\gtrsim0.1$ global and shearing sheet results differ, moreso for lower $m$.   At larger widths, global bumps are more susceptible to the RWI   than shearing sheet bumps.

In Fig. \ref{fig:global-GB-comparison}, the Rayleigh stability boundary deviates from constant $\mathcal{J}$, as described in \S\ref{sec:globalRayleigh}.  The $m=1$ mode curves to avoid the Rayleigh stability boundary,  for $W/R_\mathrm{c}\gtrsim0.1$, another manifestation of the ``halfway to Rayleigh" result.

The  $m=2,3$ modes don't share this behavior, crossing the Rayleigh boundary.  Since these modes are only weakly compressible, with $mh=k_yH=0.2,0.3$, the ``halfway to Rayleigh" behavior is not expected, and is also not seen in the comparison shearing sheet models. 

The shearing sheet stability boundary at $k_yW\simeq0.7$ is at $W/R_\mathrm{c}\simeq0.7/m\simeq0.7,.35,.23$ for the modes in Fig. \ref{fig:global-GB-comparison}. The global models are more unstable, i.e.\ to larger widths than this boundary.  This destabilizing effect diminishes for smaller $W/R_\mathrm{c}$ boundaries, as expected.

\begin{figure}\includegraphics[width=0.94\columnwidth]{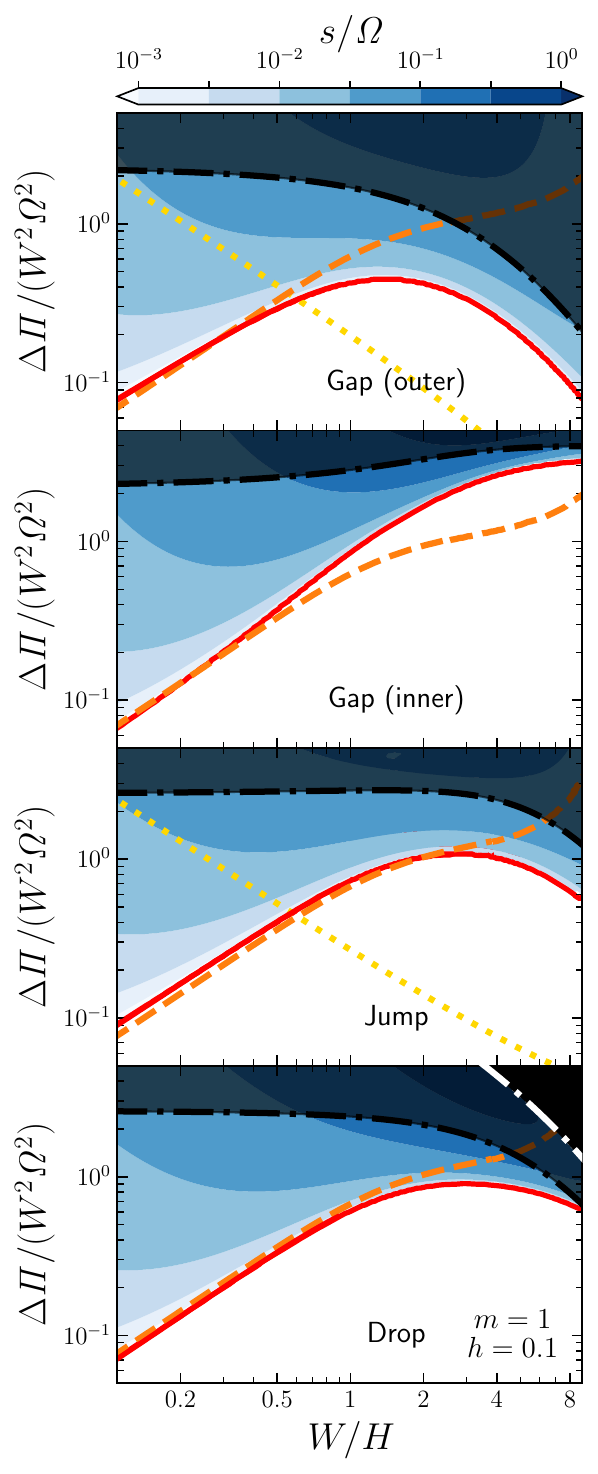}\caption{RWI growth rates of $m=1$ modes in our $h=0.1$ global disk for different shapes.  Curves for marginal stability (\emph{solid}), compressible shearing sheet marginal stability (\emph{dashed}),  Rayleigh marginal stability (\emph{black dot-dashed}) and marginal dust traps (\emph{dotted yellow})  are shown.}\label{fig:global-diff-shapes-h0.1}\end{figure}

\subsubsection{Effect of Feature Shape}

Figure \ref{fig:global-diff-shapes-h0.1} plots marginal stability curves and growth rates for $m=1,h=0.1$ RWI modes for a range of shapes (but not bumps, shown in Fig. \ref{fig:global-GB-comparison}).  
Global marginal stability is compared to shearing sheet results for $k_yH=mh=0.1$.  The agreement is again good for $W/R_\mathrm{c}\lesssim0.1$, i.e.\ $W/H\lesssim1$ on this plot.

The inner and outer gap edges show the largest differences between global and shearing sheet models, starting for $W/H\gtrsim0.5$.   The large distance between the gap center and vorticity minima ($\simeq1.7W$, see Fig.\ \ref{fig:feats}) is a natural explanation.

For most shapes, global models are more susceptible to the RWI than shearing sheet models of equivalent parameters.   Inner gap edges are the only exception (of the shapes considered).  Inner gap edges are also the only shape to show reduced Rayleigh instability at larger widths (\S\ref{sec:globalRayleigh}).  The ``halfway to Rayleigh" behavior of the RWI boundary thus also applies to  global models as they deviate from the shearing sheet approximation.

For  narrow widths, $W/H\lesssim1$, the growth rates in  Fig.\ \ref{fig:global-diff-shapes-h0.1} are consistent with shearing sheet  (cf.\ Fig.\ \ref{fig:CSS-GB} for $k_yH=0.1$),  similar for all shapes, and given approximately by Eq.\ \eqref{eq:scomp_approx}.  For wider features,  we do not offer a global correction to this analytic approximation, as the effects seem shape dependent.  The basic behavior is that growth rates steadily increase away from the RWI boundary.

\subsubsection{Halfway to Rayleigh, Globally}\label{sec:globalminkap}
We refer back to Fig.\ \ref{fig:compressible-kap2min} for the 
the minimum value of $\kappa^2/\varOmega_0^2$ along the RWI boundary for global models with $m=1,\,h=0.03$ (a slightly thinner disk than above).  The results are generally consistent with the equivalent $k_yH=mh=0.03$ shearing sheet models.  Note that global models compare to the orbital frequency $\varOmega_0(R)$, not the fixed $\varOmega$ of the shearing sheet.

For $W\gtrsim2H$, all models have the RWI boundary occurring ``halfway to Rayleigh" with $\min(\kappa^2/\varOmega_0^2)\sim0.5{-}0.6$.  The inner and outer gap edges again show the largest global corrections at larger widths.  Inner gap edges are again the most special case and the most resistant to RWI, especially for wide gaps.  Inner gap edges require the lowest values of  $\min(\kappa^2/\varOmega_0^2)$ and the largest enthalpy amplitudes (Fig.\ \ref{fig:global-diff-shapes-h0.1}) to trigger RWI.

For thinner disks, $h\lesssim0.01$, global corrections are less significant (for fixed $W/H$) and $\min(\kappa^2/\varOmega_0^2)$ more constant along the RWI boundary.  This effect is due to stronger compressibility, with $mh=k_yH<0.01$ (Fig.\ \ref{fig:effect_compressibility}). 
Disks with moderate thickness, $0.03\lesssim{h}\lesssim0.3$, are more realistic but more complicated, due to intermediate compressibility and stronger global curvature effects.

\subsection{RWI in Dust Traps}\label{sec:dusttraps}

\cite{chang23} examined which dust trapping rings became unstable to -- and would thus be modified by -- the RWI.  The condition for dust trapping is a maximum in the 
midplane pressure,
\begin{align}
P_{\rm mid} &= \varOmega_{\rm K} \frac{P_0}{c_0}  = \frac{\varOmega_{\rm K}}{\sqrt{\gamma}}\sqrt{P_0 \varSigma_0}
\end{align} 
assuming a vertically isothermal structure.

Figures \ref{fig:global-GB-comparison} and \ref{fig:global-diff-shapes-h0.1} show 
the minimum amplitude needed for dust trapping.  Note that no dust traps exist for inner gap edges or drops since they reinforce $d P_{\rm mid}/dR < 0$ instead of reversing it.  Dust traps that are stable to RWI lie in the parameter space above the yellow dotted dust trapping curves and below the solid RWI boundaries.

As in \cite{chang23}, which only considered bumps, stable (to RWI) dust traps exist above a minimum width and for a range of intermediate amplitudes.  This parameter space is larger for outer gap edges and jumps, vs.\ bumps, for reasons that can be explained by an analysis of $P_{\rm mid}$ similar to that of $\kappa^2$ in \S\ref{sec:globalRayleigh}.

We defer a more detailed study of dust trap stability that further extends \cite{chang23}.  We mainly note that such an analysis is facilitated by the insights to the RWI boundary established in this work.



\section{Conclusions}\label{sec:conclusions}

We examine the linear Rossby Wave instability (RWI) with a suite of simplified models, to gain a basic understanding of the conditions for instability, and unstable growth rates.  The disk features that trigger the RWI are best characterized by their enthalpy amplitude, $\Delta \varPi$ and width $W$.  When different combinations of temperature and density produce the same enthalpy profile, the equilibrium velocity and vorticity profiles are the same (Eqs.\ \ref{eq:v0c}, \ref{eq:epicyclic}).  We apply enthalpy features with various shapes (\S\ref{sec:shapes}) to a suite of models, in the incompressible shearing sheet (\S\ref{sec:inc_results}), the compressible shearing sheet (\S\ref{sec:comp_results}) and global models (\S\ref{sec:comp_global}).  Our main insights, explored in detail in the text, follow.

\begin{enumerate}\item The RWI in the incompressible shearing sheet (ISS) is simply characterized by two dimensionless parameters: the scaled enthalpy amplitude $\mathcal{J}=\Delta\varPi/(\varOmega W)^2$ and $k_yW$ (wavenumber times width).
The ISS RWI can be understood analytically, including the stability boundary (Eqs.\ \ref{eq:ISS-MS-fit}, \ref{eq:lith_MS}) and growth rate (Eqs.\ \ref{eq:sinc_approx}, \ref{eq:alphaint}).  The ISS RWI has a similar mechanism and eigenfunctions to the full disk RWI and to generic shear instabilities (Fig.\ \ref{fig:ISS-vort-eigfuncs-smallgr}).
 
\item The RWI in the compressible shearing sheet (CSS) requires the additional parameter, $k_yH$ (wavenumber times scale-height), an inverse Mach number.  Modes with $k_yH>1$ behave incompressibly.  Smaller $k_yH$ values show stronger compressibility effects (Fig.\ \ref{fig:effect_compressibility}).  

\item The RWI is moderately compressible in typical protoplanetary disks with aspect ratios $0.03\lesssim{h}\lesssim0.3$ \citep{cy10}.  Specifically, $m=1$ azimuthal modes have $k_yH=mh\rightarrow{h}$, and are  compressible, while $m\gtrsim1/h$ modes are incompressible.

\item The RWI is usually most readily triggered by the longest wavelength, $m=1$ modes (\citealp{ono16,chang23}; \S\ref{sec:m}).  However, in very thin disks, modes with different $m$ but $mh=k_yH\lesssim 0.01$ will have nearly the same RWI boundary, due to strong compressibility (Eq.\ \ref{eq:CSS-MS-fit}).

\item Only disk features with widths $W\lesssim 1/k_y$  can trigger the RWI (Figs.\ \ref{fig:ISS-gramp}, \ref{fig:effect_compressibility}).  In global models with a feature at radius $R_\mathrm{c}$, this limit, $W/R_\mathrm{c}\lesssim1/m$,  is relevant (i.e.\ smaller than the disk) for  $m>1$ (Fig.\ \ref{fig:global-GB-comparison} and \S\ref{sec:m}).  This  limit is roughly derived in \S\ref{sec:inc_intuitive}.

\item The RWI boundary often lies ``halfway to Rayleigh instability" in that $\min(\kappa^2)$ drops to $\sim0.5{-}0.6\,\varOmega^2$.   This behavior occurs for widths $H\lesssim{W}\lesssim1/k_y=R_\mathrm{c}/m$   (Fig. \ref{fig:compressible-kap2min}), a range that expands for thinner disks (and thus stronger compressibility, Fig.\ \ref{fig:effect_compressibility}).  This boundary is roughly derived in \S\ref{sec:int_width}.

\item For narrow features, with $W\lesssim H$ the RWI boundary follows $\Delta\varPi\propto{W}^3$ (Eq.\ \ref{eq:CSS-MS-fit}).  This scaling agrees with the low amplitude behavior in  \citet{ono16}, see Eq.\ \ref{eq:sigenth}.  We  explain the relevant factors that turn this previously known proportionality into an equality.

\item The stability boundary for the RWI of localized disk feature (with $W\lesssim0.2R_\mathrm{c}$) can be approximated by equation \eqref{eq:CSS-MS-fit}.  The enthalpy amplitude and width must be calculated (see \S\ref{sec:globalmodels}).   For wider disk features, the ``halfway to Rayleigh" criterion is a good approximation for $m=1$ modes (Figs. \ref{fig:global-GB-comparison}, \ref{fig:global-diff-shapes-h0.1}). 

\item Shape effects are generally minor when comparing the same enthalpy amplitude and width.  However bumps are the most susceptible to RWI.  Wide gaps show the largest global corrections, compared to shearing sheet models.  The inner edges of wide gaps are the least susceptible to RWI (Fig.\ \ref{fig:compressible-kap2min}, \ref{fig:global-diff-shapes-h0.1}).\end{enumerate}

This final point implies that wide, symmetric planetary gaps the outer gap edge should generally support more vigorous RWI and vortex formation. Vortices at the outer edges of gaps could be more prominent in simulations and observations for other reasons as well, including: larger area, longer orbital and viscous timescales, numerical resolution and more dust trapping \citep{fu14,lobo-gomes15,regaly17,hammer17}.  Alternately, our results imply that over longer times, the RWI should make wide planetary gaps more asymmetric, with closer and steeper inner edges.  The radial powerlaw of the background disk also affects gap asymmetries \citep{cimerman23}.

Our simplified models neglect many physical effects, notably baroclinicity, cooling, 3D motions and self-gravity.  Previous works have studied the RWI with these effects and shown their importance (\S\ref{sec:intro}).  Studying these effects with methods similar to this work---e.g.\ parameter studies of models of increasing complexity---could yield further insights.

\section*{Acknowledgements}    
We thank Leonardo Krapp, Juan Garrido-Deutelmoser, Gordon Ogilvie, Roman Rafikov, Kaitlin Kratter, Wlad Lyra, Orkan Umurhan and other participants in the  sptheory (University of Arizona Star and Planet formation Theory)  and pfits+ (Planet Formation in the Southwest) group meetings for inspiring discussions and advice.  We acknowledge support from  the NASA Theoretical and Computational Astrophysical Networks (TCAN) via grant \#80NSSC21K0497.


\bibliography{ms}{}
\bibliographystyle{aasjournal}

\end{document}